\documentclass[%
 reprint,
 amsmath,amssymb,
 aps,
pra,
]{revtex4-2}

\usepackage[colorlinks = true, linkcolor = blue, urlcolor  = blue, citecolor = blue, anchorcolor = blue]{hyperref}
\usepackage{xspace}
\usepackage{multirow, amsmath, array, booktabs, color, bm}
\usepackage{longtable,mathrsfs}
\usepackage{ epsfig, latexsym, amssymb}
\usepackage[section]{placeins}
\usepackage[textsize=footnotesize]{todonotes}
\usepackage{dcolumn}
\usepackage{physics}

\usepackage{graphicx}

\usepackage[caption=false]{subfig}
\begin{document}
\title{ Theory of the Collective Many-body Subradiance in Waveguide Quantum Electrodynamics }
\author{Xin Wang}
\affiliation{School of Physics, Sun Yat-sen University, Guangzhou 510275, China}
\author{Junjun He}
\affiliation{School of Physics, Sun Yat-sen University, Guangzhou 510275, China}

\author{Zeyang Liao}
\email[E-mail:]{liaozy7@mail.sysu.edu.cn}
\affiliation{School of Physics, Sun Yat-sen University, Guangzhou 510275, China}


\begin{abstract}

We present an analytical theory for the most subradiant modes in a finite one-dimensional emitter array coupled to either an ideal or a nonideal waveguide. Using an effective non-Hermitian Hamiltonian together with a Bragg-edge open-boundary ansatz, we derive compact expressions for the full complex collective eigenvalues, including both the linewidths and the collective energy shifts. The linewidths of the most subradiant states exhibit the characteristic $N^{-3}$ scaling in both cases, while in the deep-subwavelength regime they display even--odd oscillations due to boundary interference. In contrast, the collective energy shift approaches a separation-dependent asymptotic value with a leading finite-size correction scaling as $N^{-2}$. These results highlight the distinct physical origins of the imaginary and real parts of the subradiant eigenvalue: Bragg-edge destructive interference controls the linewidth, whereas near-field dipole--dipole interactions dominate the collective shift. Our theory provides a transparent framework beyond the ideal-waveguide limit and opens potential applications in subradiant spectroscopy and waveguide-QED-based sensing.

\end{abstract}

\maketitle

\section{Introduction}

Waveguide quantum electrodynamics (waveguide QED) has emerged as a powerful platform for studying collective light--matter interactions in reduced dimensions, where distant quantum emitters are coupled through a common photonic continuum, giving rise to long-range coherent and dissipative interactions~\cite{Kimble2008,vanLoo2013,lodahl2015,roy2017,chang2018a,sheremet2023,GonzalezTudela2024}. 
Unlike conventional free-space settings, nanophotonic waveguides reshape the electromagnetic mode structure and significantly enhance emitter--photon coupling, enabling precise control over photon transport, collective decay, and effective dipole--dipole interactions at both few- and many-emitter levels~\cite{Arcari2014,Tiecke2014,goban2014,goban2015,Volz2014,Lodahl2017,Petersen2014,Sollner2015,cheng2017,lu2025}. 
These capabilities have positioned waveguide QED as a natural test bed for exploring open many-body physics~\cite{Hood2016,Corzo2016,corzo2019,Solano2017,lalumiere2013,Sorensen2016,Fayard2021,wang2024b,jia2025}, non-Hermitian collective phenomena~\cite{nie2023,Dong2025,Yang2022,Lin2024}, topological effects~\cite{Bello2019,Nie2020,Kim2021,Pakkiam2023,lu2024,Ke2023,Xiang2025,zhu2025}, mirror-like optical responses~\cite{TsoiLaw2008,Chang2011,Chang2012,liao2016b,Mirhosseini2019,Rui2020,wang2025}, and quantum functionalities based on strongly interacting emitter arrays~\cite{gonzalez-tudela2015, gonzalez-tudela2017,liao2018,xing2024,Kannan2023,xing2024a,tian2025,xing2022}.

A central aspect of collective emission is the interplay between superradiance and subradiance \cite{GrossHaroche1982}. 
Since Dicke's seminal work~\cite{Dicke1954}, it has been recognized that interference among emitters can either dramatically enhance or suppress spontaneous emission \cite{Svidzinsky2010,liao2014,Bettles2016}. Among these effects, subradiant states are particularly intriguing because they exhibit long-lived collective excitations with strongly suppressed radiation losses, making them a key concept in cooperative quantum optics \cite{Scully2015,ke2019,masson2020,zhou2025,chu2025}, collective photon storage \cite{Kalachev2005,Shen2022,Kim2025}, and quantum sensing \cite{liao2017,wang2025a,zafrabono2025}.

Ordered emitter arrays provide a controlled setting to understand subradiance. 
In both free space and structured photonic environments, periodic arrangements support collective subradiant eigenmodes near the Brillouin-zone edge. 
In one-dimensional waveguides, these effects are further sharpened because photons propagate in restricted channels and mediate effectively infinite-range couplings~\cite{ZhengBaranger2013,zeyangliao2015,Mukhopadhyay2020,Dinc2020,Lee2025,liu2025}. 
Thus, finite atomic chains coupled to waveguides serve as an ideal model system to analyze how boundary conditions, lattice periodicity, and reservoir structure jointly determine the collective subradiance effects.

Recent studies found that subradiant states in such one-dimensional chains, either in free space or coupled to a waveguide, exhibit a characteristic finite-size behavior: the most subradiant states have decay rates $\Gamma \propto 1/N^3$ \cite{a.asenjo-garcia2017,albrecht2019,Henriet2019}. 
Zhang and M{\o}lmer theoretically proved this $N^{-3}$ scaling law~\cite{zhang2019a,Zhang2020}, clarifying the origin of extreme spectral narrowing of Bragg-edge dark modes and establishing the $N^{-3}$ law as a robust signature of one-dimensional collective subradiance within the corresponding open-chain geometry.

However, existing theories have primarily focused on the decay rate alone, largely overlooking the collective energy shift.
In addition, in realistic waveguide-QED systems, emitters may also couple to nonguided radiation modes, introducing both extra collective dissipation and coherent dipole--dipole interactions~\cite{liao2016,asenjo-garcia2017}. 
This issue becomes critical in the deep-subwavelength regime ($d \ll \lambda$), where near-field contributions scale as inverse powers of the separation. 
Although Bragg-edge interference can still efficiently suppress the radiative linewidth, the collective energy shift may remain large because it is governed by short-range near-field physics rather than by the same destructive interference that narrows the decay rate. 
Understanding realistic subradiant resonances in finite waveguide-coupled arrays thus requires analytical control of both linewidth and energy shift, with a clear separation between guided and nonguided contributions.

In this work, we systematically develop an analytical theory for the collective subradiant modes of a finite one-dimensional array of identical two-level emitters coupled simultaneously to a guided mode and to nonguided free-space vacuum modes.
The central objective is to obtain the full complex collective eigenvalue $\lambda_\xi=J_\xi-i\Gamma_\xi/2$ of the most subradiant branches, rather than only their decay rates.
Our key innovations are twofold. 
First, while previous studies focused primarily on the linewidth, we derive explicit analytical expressions for the collective energy shift of subradiant states. 
We show that after subtracting the thermodynamic limit, the energy shift scales as $N^{-2}$, revealing a qualitatively different finite-size behavior from that of the linewidth. 
Second, we uncover a striking even--odd oscillatory structure in the decay rate for nonideal waveguides in the deep-subwavelength regime: when the interatomic distance is much smaller than the resonant wavelength, the decay rate follows the characteristic $N^{-3}$ scaling and simultaneously exhibits pronounced parity-dependent oscillations. 
We derive these results analytically from first principles and confirm them by exact numerical solutions.

These results provide a unified analytical description of ultranarrow yet strongly shifted subradiant resonances in realistic waveguide-QED arrays. They clarify how Bragg-edge interference, finite-size boundary effects, and near-field dipole--dipole interactions jointly determine the collective spectrum beyond the ideal-waveguide limit. 
More broadly, our work establishes a framework for understanding subradiant spectral properties in nonideal nanophotonic systems and for evaluating their potential applications in long-range many-body physics, quantum storage, and waveguide-based quantum sensing.

The remainder of the paper is organized as follows. 
In Sec.~\ref{sec:model}, we introduce the model and the effective non-Hermitian Hamiltonian. 
In Sec.~\ref{sec:ideal-waveguide}, we derive the subradiant decay rate and energy shift in the ideal-waveguide limit. 
In Sec.~\ref{sec:nonideal-waveguide}, we derive the linewidth and collective energy shift of the most subradiant modes in the nonideal case and analyze their asymptotic behavior in the deep-subwavelength regime. 
Finally, Sec.~\ref{sec:conclusion} summarizes the main results.

\section{Effective Hamiltonian}
\label{sec:model}

\begin{figure}[!htbp]
    \centering
    {\includegraphics[width=0.9\linewidth]{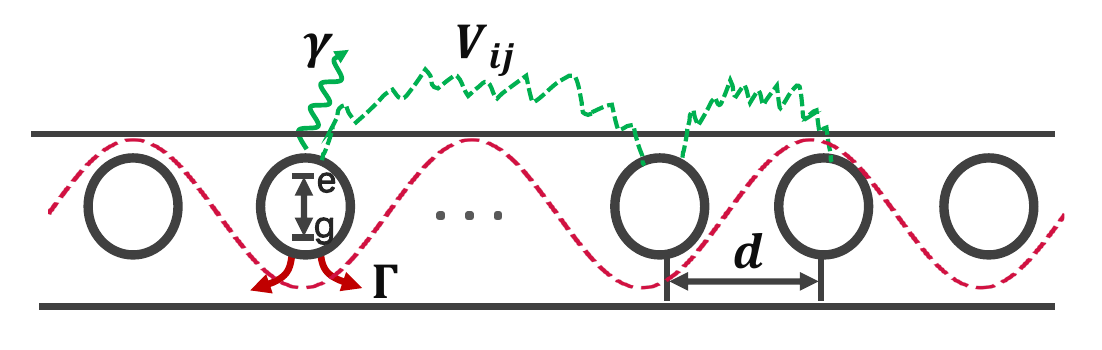}}
    \caption{
    Schematic of an array of $N$ identical two-level atoms with lattice spacing $d$ coupled to a single-mode one-dimensional waveguide. Each atom decays into the guided mode at rate $\Gamma$ and into nonguided free-space modes at rate $\gamma$. 
    }
    \label{fig:model}
\end{figure}

The theoretical model considered here is shown in Fig.~\ref{fig:model}: $N$ identical two-level atoms with transition frequency $\omega_0$ are equally spaced by $d$ along a single-mode one-dimensional waveguide. The nonguided modes are treated as free-space vacuum. After tracing out the photonic degrees of freedom under the Markov approximation, the effective Hamiltonian in the rotating frame can be written in the Green-function form~\cite{asenjo-garcia2017}
\begin{equation}
    \mathcal{H}_{\mathrm{eff}}
    =-\frac{\mu_0\omega_0^2}{\hbar}
    \sum_{j,l=1}^{N}
    \mathbf{p}_{j}^{*}\!\cdot\!
    \mathbf{G}(\mathbf{r}_j,\mathbf{r}_{l},\omega_0)
    \!\cdot\!
    \mathbf{p}_{l}\,
    \hat{\sigma}_j^{+}\hat{\sigma}_l^{-},
    \label{eq:Heff}
\end{equation}
where $\mathbf{p}_j$ and $\mathbf{p}_l$ are the transition dipole moments of the $j$th and $l$th emitters, respectively, and $\mathbf{G}(\mathbf{r},\mathbf{r}',\omega)$ is the dyadic Green function satisfying
\begin{equation}
\nabla\times\nabla\times \mathbf{G}(\mathbf{r},\mathbf{r}',\omega)
-\frac{\omega^{2}}{c^2}\epsilon(\mathbf{r},\omega)\mathbf{G}(\mathbf{r},\mathbf{r}',\omega)
=
\delta(\mathbf{r}-\mathbf{r}').
\end{equation}
Here $\mu_0$ and $\epsilon(\mathbf{r},\omega)$ are the permeability and permittivity of the environment, respectively. Throughout this work we set $\hbar=1$.

The dyadic Green function is decomposed into guided and nonguided parts,
\begin{equation}
\mathbf{G}(\mathbf{r},\mathbf{r}',\omega_0)
=
\mathbf{G}_{\mathrm{1D}}(\mathbf{r},\mathbf{r}',\omega_0)
+
\mathbf{G}_{\mathrm{NG}}(\mathbf{r},\mathbf{r}',\omega_0),
\end{equation}
where $\mathbf{G}_{\mathrm{1D}}$ describes the guided mode and $\mathbf{G}_{\mathrm{NG}}$ describes the nonguided radiation modes.

For a single-mode waveguide, the guided Green function typically has the form
$G_{\mathrm{1D}}(z,z',\omega_0)\propto e^{ik_{\mathrm{1D}}|z-z'|}$.
The corresponding guided contribution to the effective Hamiltonian is~\cite{a.asenjo-garcia2017,wang2025}
\begin{equation}
    H_{\mathrm{1D}}
    =
    -i\frac{\Gamma}{2}
    \sum_{j,l=1}^{N}
    e^{ik_{\mathrm{1D}}|z_j-z_l|}
    \hat{\sigma}_j^{+}\hat{\sigma}_l^{-},
    \label{eq:H1D}
\end{equation}
where
\begin{equation}
\Gamma
=
\frac{2\mu_0\omega_0^2}{\hbar}\,
\operatorname{Im}\!\left[
\mathbf{p}^{*}\!\cdot\!
\mathbf{G}_{\mathrm{1D}}(z_j,z_j,\omega_0)
\!\cdot\!
\mathbf{p}
\right]
\end{equation}
is the spontaneous decay rate of a single emitter into the guided mode. For $j\neq l$, Eq.~\eqref{eq:H1D} contains both the dissipative coupling
$-(i\Gamma/2)\cos(k_{\mathrm{1D}}|z_j-z_l|)$
and the coherent guided-mediated interaction
$(\Gamma/2)\sin(k_{\mathrm{1D}}|z_j-z_l|)$, where $z_{jl}=|z_j-z_l|$.
In the following sections we specialize to the resonant case $k_{\mathrm{1D}}=k_0$ in order to match the notation used in the analytical formulas.

For the nonguided electromagnetic modes, the explicit form of
\(\mathbf{G}_{\mathrm{NG}}\) depends on the detailed waveguide geometry. 
However, when the emitters are
not positioned very close to the waveguide surface and their transition frequency lies well above the waveguide
cutoff frequency, the nonguided radiation Green function can be well approximated by the free-space Green tensor, i.e., $\mathbf{G}_{\mathrm{NG}}\approx \mathbf G_0$, where \(\mathbf G_0\) is the free-space Green tensor ~\cite{LeKien2005,LeKien2017,Zhou2025Selective}. This approximation may break down when the emitters are very close to the waveguide surface or when the dielectric environment strongly reshapes the nonguided radiation continuum. In such cases, the same Green-function formalism remains valid, but \(\mathbf G_{\rm NG}\) should be replaced by the full geometry-dependent nonguided Green tensor.

Within the effective free-space nonguided-channel approximation, the corresponding Hamiltonian is
\begin{equation}
    H_{\mathrm{fs}}
    =
    -\frac{i\gamma}{2}\sum_{j=1}^N \hat{\sigma}_j^+\hat{\sigma}_j^-
    -i\sum_{j\neq l}
    V_{jl}\,e^{ik_0 z_{jl}}
    \hat{\sigma}_j^{+}\hat{\sigma}_l^{-},
    \label{eq:Hng}
\end{equation}
where $\gamma$ is the spontaneous decay rate of a single atom in free space, and \cite{liao2016,liao2017}
\begin{equation}
V_{jl}
=
\frac{3\gamma}{4}
\left[
-\frac{i}{k_0 z_{jl}}
+
\frac{1}{(k_0 z_{jl})^{2}}
+
\frac{i}{(k_0 z_{jl})^{3}}
\right],
\label{eq:Vjl-perp}
\end{equation}
where we assume that the transition dipole moments of the atoms are perpendicular to the atom-chain direction.
Only the geometry-dependent interaction part is kept explicitly in the free-space contribution; the diagonal geometry-independent Lamb shift is absorbed into the renormalized atomic transition frequency.

The total non-Hermitian effective Hamiltonian is therefore
\begin{equation}
    \mathcal{H}_{\mathrm{eff}}
    =
    H_{\mathrm{1D}}+H_{\mathrm{fs}}.
    \label{eq:Hefftotol}
\end{equation}
By solving the eigenproblem
\begin{equation}
  \mathcal{H}_{\mathrm{eff}}\ket{\phi_\xi}
  =
  \lambda_\xi\ket{\phi_\xi},
\end{equation}
we can obtain the effective eigenenergy 
\begin{equation}
\lambda_\xi = J_\xi - \frac{i}{2}\Gamma_\xi,
\end{equation}
where $\Gamma_\xi \equiv -2\,\operatorname{Im}(\lambda_\xi)$ is the collective linewidth and $J_\xi \equiv \operatorname{Re}(\lambda_\xi) $ is the collective energy shift. In the single-excitation subspace, the collective eigenstates can be written as
\begin{equation}
  \ket{\phi_\xi}
  =
  \sum_{j=1}^{N}c_\xi(j)\ket{e_j},
  \qquad
  \sum_{j=1}^{N}|c_\xi(j)|^2=1,
\end{equation}
where $\ket{e_j}$ denotes the state with only the $j$th atom excited.

In the following, we analyze the linewidths and collective energy shifts of the collective subradiant modes in the case of ideal and nonideal waveguides.

\section{Ideal waveguide}
\label{sec:ideal-waveguide}

We first consider the ideal-waveguide limit, in which decay into
nonguided modes is neglected, i.e., $\gamma\rightarrow0$. The effective
Hamiltonian is then given by Eq.~\eqref{eq:H1D}. Following the approach
of Zhang and M{\o}lmer~\cite{zhang2019a}, a collective eigenstate of a
finite open chain is constructed as a superposition of two degenerate
Bloch states,
\begin{equation}
|\phi_k\rangle=A|k\rangle+B|-k\rangle,
\qquad
|k\rangle=
\frac{1}{\sqrt{N}}
\sum_{\ell=1}^{N}e^{ikz_\ell}|e_\ell\rangle .
\label{eq:ideal_bloch_superposition_main}
\end{equation}
A single Bloch state is generally not an eigenstate of the finite chain
because the open boundaries generate residual tails proportional to
the two superradiant states $|\pm k_0\rangle$. Requiring these tails to cancel selects a discrete set of complex wave numbers. 
After sorting all eigenvalues with decay rates increasing from $ \xi=1 $ to $ \xi=N $, for $\xi\ll N$,
the two families of most subradiant solutions are located near the
center and the edge of the first Brillouin zone and satisfy
\begin{equation}
\begin{aligned}
k_{\xi}^{(0)}d
&\approx
\frac{\pi\xi}{N}
\left[
1-\frac{i}{N}
\cot\!\left(\frac{k_0d}{2}\right)
\right],
&& k_{\xi}^{(0)}\approx0,
\\
\left(k_{\xi}^{(\pi)}+\frac{\pi}{d}\right)d
&\approx
\frac{\pi\xi}{N}
\left[
1+\frac{i}{N}
\tan\!\left(\frac{k_0d}{2}\right)
\right],
&& k_{\xi}^{(\pi)}\approx-\frac{\pi}{d}.
\end{aligned}
\label{eq:ideal_two_k_branches_main}
\end{equation}
The imaginary corrections in
Eq.~\eqref{eq:ideal_two_k_branches_main} are of order $N^{-2}$ and
generate the subradiant linewidths. The corresponding exact
Bloch-state dispersion is
\begin{equation}
\omega_k^{(\mathrm{1D})}
=
\frac{\Gamma}{2}
\frac{\sin(k_0d)}
{\cos(kd)-\cos(k_0d)} .
\label{eq:omega1D-main}
\end{equation}
Writing
\begin{equation}
\omega_\xi^{(\mathrm{1D})}
=
J_\xi-\frac{i}{2}\Gamma_\xi ,
\label{eq:ideal_complex_eigenvalue_main}
\end{equation}
and substituting the two solutions in
Eq.~\eqref{eq:ideal_two_k_branches_main} into
Eq.~\eqref{eq:omega1D-main}, we obtain the two asymptotic subradiant
families derived explicitly in
Appendix~\ref{app:ideal-waveguide-derivation}.

\begin{figure*}[!ht]
    \centering
    \includegraphics[width=0.9\textwidth]{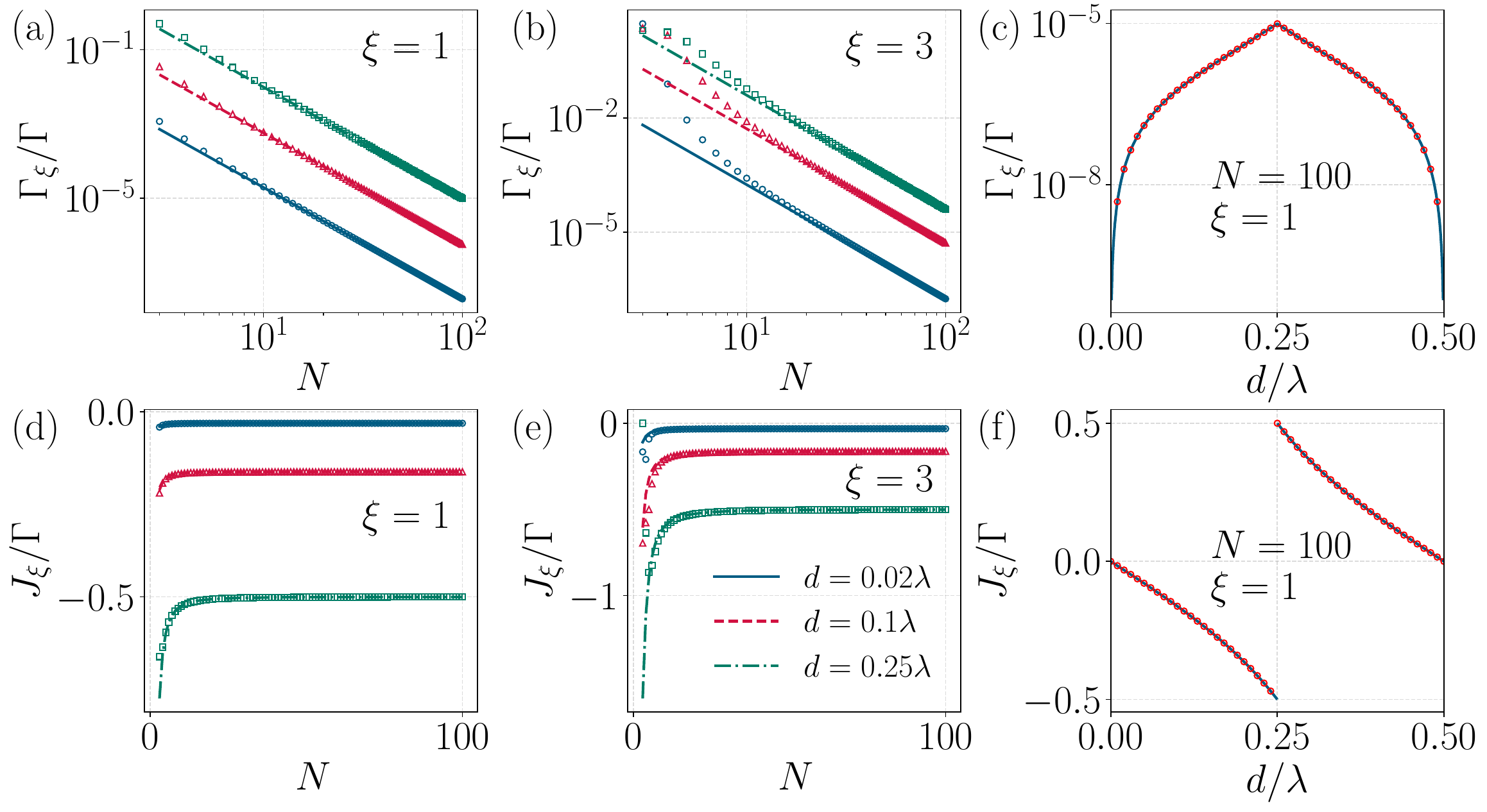}
    \caption{
    Benchmark of the ideal-waveguide asymptotic formulas against exact
    numerical diagonalization.
    (a),(b) Collective linewidths $\Gamma_\xi/\Gamma$ versus atom
    number $N$ for $\xi=1$ and $\xi=3$, respectively, at three lattice
    spacings: $d=0.02\lambda$ (solid blue lines and open circles),
    $d=0.1\lambda$ (dashed red lines and open triangles), and
    $d=0.25\lambda$ (dash-dotted green lines and open squares).
    (c) Linewidth $\Gamma_{\xi=1}/\Gamma$ versus $d/\lambda$ at fixed
    $N=100$. The analytical curve follows the
    $k\simeq-\pi/d$ family for $0<d<\lambda/4$ and the $k\simeq0$
    family for $\lambda/4<d<\lambda/2$. The two linewidth expressions
    coincide at $d=\lambda/4$.
    (d),(e) Collective energy shifts $J_\xi/\Gamma$ versus $N$ for
    $\xi=1$ and $\xi=3$, respectively, at the same three spacings.
    For $d=0.25\lambda$, the negative-shift $k\simeq-\pi/d$ member of
    the linewidth-degenerate pair is shown.
    (f) Collective energy shift $J_{\xi=1}/\Gamma$ versus $d/\lambda$
    at fixed $N=100$. At $d=\lambda/4$, the two modes have the same
    linewidth but opposite collective shifts; both branch values are
    shown, and the linewidth-selected shift is therefore not uniquely
    defined at the crossing.
    In all panels, lines denote the analytical results, while open
    symbols denote the exact numerical eigenvalues of the effective
    Hamiltonian. The linewidth data are compared with
    Eq.~\eqref{eq:Gamma_zhang_main}, and the collective-shift data are
    compared with Eq.~\eqref{eq:J_ideal_main_equiv}.
    }
    \label{fig:ideal_summary}
\end{figure*}

Because the ideal-waveguide spectrum is periodic in $d$ with period
$\lambda/2$, it is sufficient to consider $0<d<\lambda/2$. In this
interval, the $k_{\xi}^{(\pi)}\approx-\pi/d$ family has the smaller
linewidth for $0<d<\lambda/4$, whereas the
$k_{\xi}^{(0)}\approx0$ family has the smaller linewidth for
$\lambda/4<d<\lambda/2$. Therefore, the linewidth of the narrowest
subradiant mode is
\begin{equation}
\Gamma_\xi(N,d)
\approx
\frac{\Gamma}{2}
\frac{\pi^2\xi^2}{N^3}
\begin{cases}
\displaystyle
\frac{\sin^2(k_0d/2)}
{\cos^4(k_0d/2)},
&
0<d<\dfrac{\lambda}{4},
\\[1.0em]
\displaystyle
\frac{\cos^2(k_0d/2)}
{\sin^4(k_0d/2)},
&
\dfrac{\lambda}{4}<d<\dfrac{\lambda}{2}.
\end{cases}
\label{eq:Gamma_zhang_main}
\end{equation}
Both families retain the characteristic
$\Gamma_\xi\propto\xi^2/N^3$ scaling. The collective energy shift of
the linewidth-selected mode is
\begin{widetext}
\begin{equation}
J_\xi(N,d)
\approx
\begin{cases}
\begin{aligned}
&-\frac{\Gamma}{2}
\tan\!\left(\frac{k_0d}{2}\right)
-\frac{\Gamma}{8}
\left(\frac{\pi\xi}{N}\right)^2
\frac{\sin(k_0d/2)}
{\cos^3(k_0d/2)}
\end{aligned}
&
0<d<\dfrac{\lambda}{4},
\\[1.4em]
\begin{aligned}
&\frac{\Gamma}{2}
\cot\!\left(\frac{k_0d}{2}\right)
+\frac{\Gamma}{8}
\left(\frac{\pi\xi}{N}\right)^2
\frac{\cos(k_0d/2)}
{\sin^3(k_0d/2)}
\end{aligned}
&
\dfrac{\lambda}{4}<d<\dfrac{\lambda}{2}.
\end{cases}
\label{eq:J_ideal_main_equiv}
\end{equation}
\end{widetext}
At $d=\lambda/4$, the two families have the same linewidth,
\begin{equation}
\Gamma_\xi^{(0)}
=
\Gamma_\xi^{(\pi)}
\approx
\Gamma\frac{\pi^2\xi^2}{N^3},
\end{equation}
but opposite collective shifts,
\begin{equation}
J_\xi^{(0)}
=-J_\xi^{(\pi)}\approx 
\frac{\Gamma}{2}
\left[
1+\frac{\pi^2\xi^2}{2N^2}
\right],
\label{eq:J_switching_point_main}
\end{equation}
Thus the linewidth remains continuous at the switching point, while
the energy shift of the mode selected solely by the smallest linewidth
is not single valued there.

Figure~\ref{fig:ideal_summary} compares
Eqs.~\eqref{eq:Gamma_zhang_main} and
\eqref{eq:J_ideal_main_equiv} with exact numerical diagonalization.
The analytical expressions reproduce both the $\xi^2/N^3$ linewidth
scaling and the switch of the narrowest mode at $d=\lambda/4$. Away
from the switching point, the collective shift approaches an
$N$-independent band-edge value, with a finite-size correction
proportional to $\xi^2/N^2$. The asymptotic agreement improves for
smaller $\xi/N$, as expected from
Eq.~\eqref{eq:ideal_two_k_branches_main}.

For the collective linewidth,
Figs.~\ref{fig:ideal_summary}(a) and
\ref{fig:ideal_summary}(b) show $\Gamma_\xi/\Gamma$ as a function of
$N$ for $\xi=1$ and $\xi=3$, respectively, at three representative
spacings. In all cases, the analytical results reproduce the
characteristic subradiant scaling
$\Gamma_\xi\propto\xi^2/N^3$. The agreement improves with increasing
$N$ and is better for the lower-index mode $\xi=1$, consistently with
the condition $\xi\ll N$ underlying the Bragg-edge expansion. The
larger deviations at small $N$, particularly for $\xi=3$, originate
from higher-order finite-size corrections that are neglected in the
asymptotic formula.
From Fig.~\ref{fig:ideal_summary}(c), we can see that the narrowest mode switches between
the two Bragg-edge families at $d=\lambda/4$. Its linewidth increases
for $0<d<\lambda/4$, decreases for
$\lambda/4<d<\lambda/2$, and remains continuous at the switching
point to the asymptotic order retained here.

The collective energy shift shows a markedly different behavior.
As seen in Fig.~\ref{fig:ideal_summary}(d)--\ref{fig:ideal_summary}(f), the analytical formula in Eq.~\eqref{eq:J_ideal_main_equiv} captures both the $N$ and $d$ dependence well.
For fixed spacing, $J_\xi$ rapidly approaches an $N$-independent band-edge value, while the residual mode-index dependence appears only through the finite-size correction proportional to $\xi^2/N^2$.
In particular, for $N=100$ and $\xi=1$, the system exhibits a distinct jump at $d=\lambda/4$, arising from the switch between the negative- and positive-shift branches, as seen in Fig.~\ref{fig:ideal_summary}(f)

Once the array enters the deep-subwavelength regime in a nonideal
waveguide, the nonguided free-space interaction becomes
essential. In that case, both the linewidth and the collective energy
shift must be rederived from the total effective Hamiltonian in
Eq.~\eqref{eq:Hefftotol}, since the free-space near-field contribution
can dominate the ideal-waveguide result.

\section{Nonideal waveguide}
\label{sec:nonideal-waveguide}

In contrast to the ideal-waveguide problem, the two asymptotic
subradiant families are no longer equivalent when coupling to
nonguided free-space modes is included. In the deep-subwavelength
regime considered here, the modes near $k\simeq0$ lie inside the
free-space light cone and generally acquire substantial radiative
loss. By contrast, the Brillouin-zone-edge modes satisfy
$|k_b|\simeq\pi/d\gg k_0$ and lie far outside the free-space light
cone. They therefore constitute the most subradiant family of the
nonideal system.

For open boundary conditions, these Bragg-edge modes can be
approximated by
\begin{equation}
c_\xi(j)
\approx
\sqrt{\frac{2}{N+1}}
\sin\!\left(\frac{\pi\xi j}{N+1}\right)e^{ik_bz_j},
\label{eq:Dirichlet_main}
\end{equation}
where $\xi=1,2,\ldots$, $\xi\ll N$, and
$k_b\simeq\pm\pi/d$. The two signs represent the same
Brillouin-zone edge and are related by inversion symmetry; we take
$k_b\simeq\pi/d$ in the following. Thus, the one-family treatment
below is restricted to the most subradiant sector in the
deep-subwavelength regime $k_0d\ll1$ and is not intended to describe
the full nonideal spectrum at arbitrary spacing.

The total linewidth can be decomposed as
\begin{equation}
\Gamma_\xi = \Gamma_\xi^{(\mathrm{1D})}+\Gamma_\xi^{(\mathrm{fs})}
 \label{eq:Gammatotal_main}
\end{equation}
where
\begin{equation}
  \Gamma_\xi^{(\mathrm{1D})}
  =\Gamma\left|\sum_{j=1}^{N}c_\xi(j)e^{ik_0 z_j}\right|^2
  \label{eq:Gamma1D_main}
\end{equation}
is the decay rate due to the waveguide mode and
\begin{equation}
  \Gamma_\xi^{(\mathrm{fs})}
  =\gamma\sum_{j,l=1}^{N}c_\xi^\ast(j)c_\xi(l)
  \mathcal{K}_{\mathrm{fs}}\!\big(k_0 z_{jl}\big)
  \label{eq:Gammafs_main}
\end{equation}
is the decay due to nonguided free space modes. Here, the free-space kernel is defined by
$ 2\,\operatorname{Re}\big[V(r)e^{ik_0 z}\big]
  \equiv \gamma\,\mathcal{K}_{\rm fs}(k_0 z)
$
with 
\begin{equation}
  \mathcal{K}_{\rm fs}(x)
  = \frac{3}{2}\left[
      \frac{\sin x}{x}
      +\frac{\cos x}{x^2}
      -\frac{\sin x}{x^3}
    \right].
  \label{eq:Kfs}
\end{equation}
Here the diagonal single-atom free-space decay has been included in the compact
kernel form by taking the regularized value $\mathcal{K}_{\mathrm{fs}}(0)=1$.

Likewise, the collective energy shift can be decomposed as
\begin{equation}
J_\xi=J_\xi^{(\mathrm{1D})}+J_\xi^{(\mathrm{fs})},
 \label{eq:Jtotal_main}
\end{equation}
where
\begin{equation}
  J_\xi^{(\mathrm{1D})}
  =\frac{\Gamma}{2}\sum_{j\neq l}^{N}c_\xi^\ast(j)c_\xi(l)\sin(k_0 z_{jl}),
  \label{eq:J1D_main}
\end{equation}
is the energy shift due to interaction with the waveguide mode, and 
\begin{equation}
  J_\xi^{(\mathrm{fs})}
  = \frac{\gamma}{2}\sum_{j\neq l}^{N}c_\xi^\ast(j)c_\xi(l)
  \mathcal{L}_{\mathrm{fs}}(k_0 z_{jl}).
  \label{eq:Jfs_main}
\end{equation}
is the shift due to the nonguided free-space modes. 

Only the geometry-dependent collective energy shift is retained in the free-space part, while the diagonal self-energy in the free-space part is geometry independent and absorbed into the renormalized atomic resonance.
The imaginary part of $V(r)e^{i k_0 z}$ defines the shift kernel
$ 2\Im\!\left[V(r)e^{i k_0 z}\right]
  \equiv
  \gamma\mathcal{L}_{\mathrm{fs}}(k_0z)$
with
\begin{equation}
  \mathcal{L}_{\mathrm{fs}}(x)
  =\frac{3}{2}
  \left[
    -\frac{\cos x}{x}
    +\frac{\sin x}{x^2}
    +\frac{\cos x}{x^3}
  \right].
  \label{eq:Lfs_main}
\end{equation}

\subsection{Imaginary part: decay rate of the most subradiant modes}
\label{sec:linewidth}

We now derive compact analytical expressions for the imaginary part of the eigenvalue of the most subradiant modes in the deep-subwavelength regime.

\subsubsection{Calculation of $\Gamma_\xi^{(\mathrm{1D})}$}

The decay rate due to the guided-mode is given by Eq. \eqref{eq:Gamma1D_main} which can be rewritten as
\begin{equation}
  \Gamma_\xi^{(\mathrm{1D})}
  = \Gamma\,|S_\xi|^2,\quad
  S_\xi = \sum_{j=1}^N c_\xi(j)\,e^{ik_{0} z_j}.
   \label{eq:gammaxi}
\end{equation}

For the Bragg-edge branch, $ e^{ik_b z_j}\simeq(-1)^{j-1} $ when $ k_b\simeq \pi/d $. Using Eqs. \eqref{eq:Dirichlet_main}  and \eqref{eq:gammaxi}, we have
\begin{equation}
\begin{aligned}
  S_\xi
  &= \sqrt{\frac{2}{N+1}}\sum_{j=1}^N(-1)^{j-1}
      \sin(aj)\,e^{i\beta(j-1)}  \\
 & = i\sqrt{\frac{2}{N+1}}\frac{(1-e^{2ia})[1+(-1)^{N+\xi}e^{i(N+1)\beta}]}{2(e^{ia}+e^{i\beta})(1+e^{i(a+\beta)})}.
      \end{aligned}
      \label{eq:sxi}
\end{equation}
where $\beta\equiv k_{0}d$ and $a=\pi\xi/(N+1)$. For most subradiant modes $ \xi\ll N $, $ a\ll1 $ .

Now, we consider the deep-subwavelength regime, $\beta \ll 1$, this reduces to
\begin{equation}
  S_\xi\approx a\sqrt{\frac{2}{N+1}}\frac{1+(-1)^{N+\xi}e^{i(N+1)\beta}}{4}.
      \label{eq:sxi1}
\end{equation}
Inserting Eq. \eqref{eq:sxi1} into Eq. \eqref{eq:gammaxi} yields
\begin{equation}
  \begin{aligned}
      \Gamma_\xi^{(\mathrm{1D})}(N,d) 
  \approx
  &\frac{\pi^2\xi^2}{(N+1)^3}\,
  \frac{\Gamma}{4}
  \Big[1+(-1)^{N+\xi}\cos\big((N+1)k_0d\big)\Big]. \\
  \end{aligned}
  \label{eq:Gamma-1D-final}
\end{equation}
This is the decay rate of the most subradiant states due to the guided modes in the deep-subwavelength regime. $ \xi=1 $ corresponds to the most subradiant state.

\subsubsection{Calculation of $\Gamma_\xi^{({\rm fs})}$}
\label{subsec:fs-linewidth}

Now we derive the free-space contribution to the decay rate of the Bragg-edge subradiant modes which is given by Eq. \eqref{eq:Gammafs_main}.

The free-space kernel $\mathcal{K}_{\rm fs}\!\big(k_0 z_{jl}\big)$ admits the explicit form shown in Eq.~\eqref{eq:Kfs}. 
Alternatively, the free-space decay can also be viewed as an angular average over all outgoing plane-wave modes. For a chain oriented along the $z$ axis with atomic dipoles perpendicular to the chain direction, use the angular-average representation of the kernel (derived in Appendix~\ref{app:Kfs})
\begin{equation}
  \mathcal{K}_{\rm fs}(x)
  = \frac{3}{8}\int_{-1}^{1}\!\mathrm{d}\mu\,(1+\mu^2)\,e^{i x\mu},
  \qquad
  \mu\equiv\cos\theta.
  \label{eq:Kfs-angular}
\end{equation}
Physically, $\mu=\cos\theta$ parametrizes the angle between the outgoing
photon's wave vector and the chain axis, while $(1+\mu^2)$ encodes the angular dependence of the dipole radiation pattern (summed over polarizations).
Equation~\eqref{eq:Kfs-angular} therefore describes the integral over all single-photon radiation channels in free space.

Because the integration range is symmetric and the weight $(1+\mu^2)$ is even in $\mu$, the odd sine part vanishes. Therefore, inside the angular integral, $e^{ik_0\mu |z_j-z_l|}$ is equivalent to $e^{ik_0\mu (z_l-z_j)}$ after integration over $\mu\in[-1,1]$. 
Substituting Eq.~\eqref{eq:Kfs-angular} into Eq.~\eqref{eq:Gammafs_main} thus gives
\begin{equation}
\begin{aligned}
  \Gamma_\xi^{(\mathrm{fs})}
  &= \frac{3\gamma}{8}
      \int_{-1}^{1}\!\mathrm{d}\mu\,(1+\mu^2)
      \sum_{j,l=1}^N
      c_\xi^\ast(j)c_\xi(l)\,
      e^{i k_0\mu(z_l-z_j)}.
\end{aligned}
\end{equation}
Introducing the mode-dependent structure factor
\begin{equation}
  S_\xi(\kappa)
  \equiv \sum_{j=1}^N c_\xi(j)\,e^{i\kappa z_j},
  \qquad
  \kappa\equiv k_0\mu,
\end{equation}
we obtain
\begin{equation}
  \sum_{j,l=1}^N
      c_\xi^\ast(j)c_\xi(l)\,
      e^{i k_0\mu(z_l-z_j)}
  = \Big|S_\xi(\kappa)\Big|^2.
\end{equation}
Hence
\begin{equation}
\begin{aligned}
  \Gamma_\xi^{(\mathrm{fs})}
  &= \frac{3\gamma}{8}\int_{-1}^{1}\!\mathrm{d}\mu\,(1+\mu^2)\,
      \big|S_\xi(k_0\mu)\big|^2  \\
  &= \frac{3\gamma}{4}\int_{0}^{1}\!\mathrm{d}\mu\,(1+\mu^2)\,
      \big|S_\xi(k_0\mu)\big|^2.
  \label{eq:Gamma-fs-Sxi}
\end{aligned}
\end{equation}
This shows that the free-space linewidth is an angular average of the
mode-dependent structure factor, weighted by $(1+\mu^2)$.

For a uniformly spaced chain with $z_j=(j-1)d$, the structure factor becomes
\begin{equation}
  S_\xi(\kappa)
  = \sum_{j=1}^N c_\xi(j)\,e^{i\kappa (j-1)d},
  \qquad
  \kappa = k_0\mu.
\end{equation}
This has the same algebraic form as in the waveguide contribution, 
except that the effective longitudinal wave vector is continuously sampled inside the free-space light cone,

Using the Bragg-edge OBC mode introduced above [Eq.~\eqref{eq:Dirichlet_main}], the dominant lattice momentum is centered near the Brillouin-zone edge
$k_b\simeq \pi/d$. In the deep-subwavelength regime $k_0 d\ll1$, the radiative interval $|\tilde\kappa \equiv \kappa d|\le k_0d$ lies far from the Bragg point $\pi$.
Therefore the bulk Bragg oscillation is outside the free-space light cone and cannot radiate at leading order. The remaining linewidth arises from the finite boundaries of the chain, namely the interference between the two ends.

By the same algebra that leads to Eq.~\eqref{eq:sxi} for the waveguide contribution, one obtains the corresponding closed form here after the replacement $\beta \to \tilde\kappa$.
In the deep-subwavelength limit $\tilde\kappa\ll1$, and for most subradiant modes $\xi\ll N$, keeping the leading nontrivial order in $a$ gives
\begin{equation}
  S_\xi(\kappa)
  \approx a\sqrt{\frac{2}{N+1}}\,
  \frac{1+(-1)^{N+\xi}e^{i(N+1)\tilde\kappa}}{4}
  +\mathcal{O}(a^2).
  \label{eq:Sxi-smallkappa}
\end{equation}
Therefore,
\begin{equation}
\begin{aligned}
  \big|S_\xi(\kappa)\big|^2
  &\approx
  \Big|
  a\sqrt{\frac{2}{N+1}}
  \frac{1+(-1)^{N+\xi}e^{i(N+1)\tilde\kappa}}{4}
  \Big|^2
  +\mathcal{O}(a^4) \\
  &=
  \frac{a^2}{4(N+1)}
  \Big[1+(-1)^{N+\xi}\cos\big((N+1)\tilde\kappa\big)\Big]
  +\mathcal{O}(a^4).
\end{aligned}
  \label{eq:Sxi2-smallkappa}
\end{equation}
Substituting $a=\pi\xi/(N+1)$, we arrive at
\begin{equation}
  \big|S_\xi(\kappa)\big|^2
  \approx
  \frac{\pi^2\xi^2}{4(N+1)^3}\,
  \Big[1+(-1)^{N+\xi}\cos\big((N+1)\tilde\kappa\big)\Big]
  +\mathcal{O}(a^4).
  \label{eq:Sxi2-final}
\end{equation}

Substituting Eq.~\eqref{eq:Sxi2-final} into Eq.~\eqref{eq:Gamma-fs-Sxi} and defining $\theta_{N+1} \equiv (N+1)k_0 d,$
we obtain
\begin{equation}
\Gamma_\xi^{(\mathrm{fs})}
\approx
\frac{\pi^2\xi^2}{(N+1)^3}\,
\frac{3\gamma}{16}\,
\mathcal{I}_\xi(\theta_{N+1}),
\label{eq:Gammafs_step1}
\end{equation}
with
\begin{equation}
\mathcal{I}_\xi(\theta)
\equiv
\int_0^1 d\mu\,(1+\mu^2)
\Big[1+(-1)^{N+\xi}\cos(\theta\mu)\Big].
\label{eq:Itheta_def}
\end{equation}
The integral naturally separates into a constant background term and a
boundary-induced interference term,
\begin{equation}
\mathcal{I}_\xi(\theta)
=
\int_0^1 d\mu\,(1+\mu^2)
+
(-1)^{N+\xi}
\int_0^1 d\mu\,(1+\mu^2)\cos(\theta\mu).
\label{eq:Itheta_split}
\end{equation}

Evaluating the angular integrals (Appendix~\ref{app:angular_integral}) yields
\begin{equation}
  \Gamma_\xi^{(\mathrm{fs})}(N,d)
  \approx
  \frac{\pi^2\xi^2}{(N+1)^3}\,
  \frac{\gamma}{4}
  \Big[1+(-1)^{N+\xi}\mathcal{K}_{\rm fs}(\theta_{N+1})\Big].
  \label{eq:Gamma-fs-analytic}
\end{equation}
Equation~\eqref{eq:Gamma-fs-analytic} shows that the free-space contribution preserves the same $(N+1)^{-3}$ scaling as the guided-mode part.
The parity-dependent oscillatory factor is the boundary interference that survives after the light-cone selection rule suppresses radiation from the bulk Bragg momentum.

To confirm the validity of the above derivation, we numerically evaluate Eq. \eqref{eq:Gammafs_main} which can be written as
\begin{equation}
\Gamma_{\xi}^{(\mathrm{fs})}=\gamma\mathcal{K}_{\rm fs}(0)\mathcal{C}_\xi(0)+2\gamma\sum_{\Delta=1}^{N-1}\mathcal{K}_{\rm fs}(\beta\Delta)\mathcal{C}_\xi(\Delta) 
\label{eq:Gammafs-discrete}
\end{equation}
where 
\begin{equation}
\mathcal{C}_\xi(\Delta)=\sum_{n=1}^{N-\Delta}c_\xi^{*}(n+\Delta)c_\xi(n)
\label{eq:autocorrelation}
\end{equation}
is the autocorrelation function. It can be shown that  (see Appendix~\ref{app:S} for the detailed derivation) 
\begin{equation}
\mathcal{C}_\xi(\Delta)=\frac{(-1)^\Delta}{N+1}
   \Big[(N+1-\Delta)\cos(a\Delta)+\cot a\,\sin(a\Delta)\Big].
\end{equation}

To explicitly expose the $(N+1)^{-3}$ scaling, we can rewrite $\mathcal{C}_\xi(\Delta)$ as
\begin{equation}
 \mathcal{C}_\xi(\Delta)
  = (-1)^\Delta\,\frac{\pi^2\xi^2}{(N+1)^3}\,
    W_{N+1}(\Delta;\xi),
    \label{eq:autocorrelation-scaled}
\end{equation}
where
\begin{equation}
  W_{N+1}(\Delta;\xi)
  := \frac{(N+1-\Delta)\cos(a\Delta)
           +\cot a\,\sin(a\Delta)}{a^2}.
\end{equation}
Inserting Eq. \eqref{eq:autocorrelation-scaled} into Eq.~\eqref{eq:Gammafs-discrete} we can obtain
\begin{equation}
  \Gamma_\xi^{({\rm fs})}(N,d)
  = \gamma\, \frac{\pi^2\xi^2}{(N+1)^3}\,
      \mathcal{F}_{N+1}^{({\rm fs})}(\beta;\xi),
  \label{eq:Gamma-fs-env}
\end{equation}
where the dimensionless prefactor is defined by
\begin{equation}
  \mathcal{F}_{N+1}^{({\rm fs})}(\beta;\xi)
  := \frac{N+1}{a^2}+2\sum_{\Delta=1}^{N-1}
      (-1)^\Delta W_{N+1}(\Delta;\xi)\,
      \mathcal{K}_{\rm fs}(\beta\Delta).
  \label{eq:F-def}
\end{equation}

Equation~\eqref{eq:Gamma-fs-env} is exact within the Bragg-edge Dirichlet-mode ansatz which can be numerically calculated, while Eq. \eqref{eq:Gamma-fs-analytic} provides a compact analytic expression in the deep-subwavelength and small-$\xi$ limit. To confirm the validity of Eq. \eqref{eq:Gamma-fs-analytic}, we can numerically compare the results between Eqs. \eqref{eq:Gamma-fs-analytic} and \eqref{eq:Gamma-fs-env}.
From Eq.~\eqref{eq:Gamma-fs-analytic} we can also read off the corresponding
dimensionless prefactor in the deep-subwavelength regime,
\begin{equation}
 \mathcal{F}_{N+1,\text{ana}}^{({\rm fs})}(\beta;\xi)
  \approx \frac{1}{4}\Big[1+(-1)^{N+\xi}\mathcal{K}_{\rm fs}((N+1)\beta)\Big],
  \label{eq:FN-fs-analytic}
\end{equation}

\begin{figure*}[!htbp]
    \centering
    \includegraphics[width=0.9\linewidth]{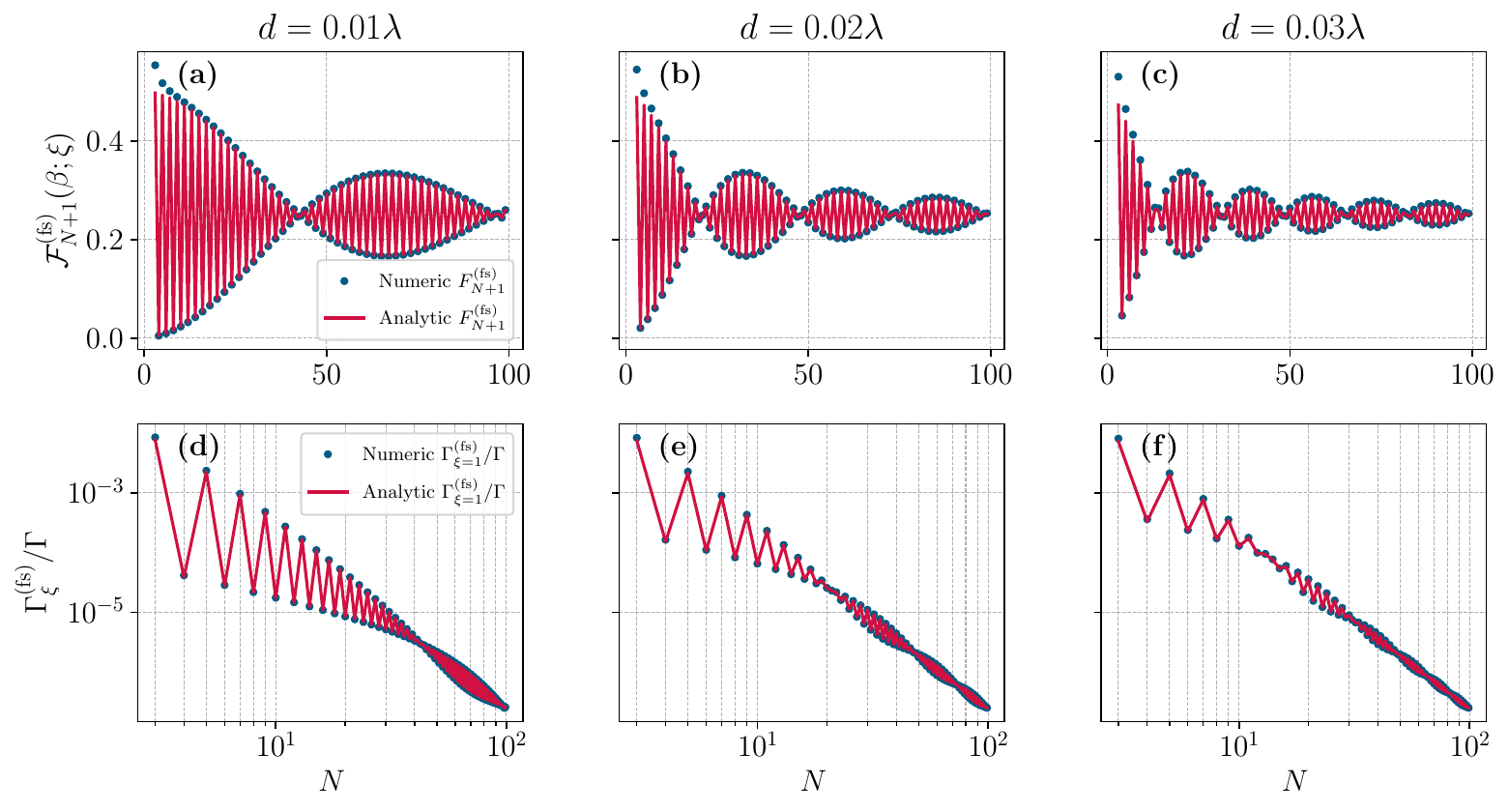}
    \caption{Comparison between the numerical evaluation and the analytic formulas for the free-space decay rate.
    (a)–(c): Dimensionless prefactor $F^{(\mathrm{fs})}_{N+1}(\beta;\xi)$ as a function of $N$ for $d=0.01\lambda$, $0.02\lambda$, and $0.03\lambda$ with $\xi=1$, obtained from the discrete sum [Eq.~\eqref{eq:F-def}] (symbols) and the analytic expression [Eq.~\eqref{eq:FN-fs-analytic}] (solid lines). 
(d)–(f): Corresponding free-space linewidth $\Gamma^{(\mathrm{fs})}_\xi(N,d)$ computed from Eq.~\eqref{eq:Gamma-fs-analytic} (solid lines) and Eq.~\eqref{eq:Gamma-fs-env} (symbols). In all panels, $\gamma=0.1\Gamma$.
    }
    \label{fig:fs_discrete}
\end{figure*}

We evaluate
$\mathcal{F}_{N+1}^{({\rm fs})}(\beta;\xi)$ numerically from the discrete sum
\eqref{eq:F-def} and compare it with the analytic prediction
\eqref{eq:FN-fs-analytic}, as well as with the resulting
$\Gamma_\xi^{({\rm fs})}(N,d)$ from Eq.~\eqref{eq:Gamma-fs-env}. The results are shown in Fig.~\ref{fig:fs_discrete} for several deep-subwavelength separations $d\ll\lambda$ and for the most subradiant mode $\xi=1$, as an explicit example. From Figs. ~\ref{fig:fs_discrete}(a-c), we can see that $ \mathcal{F}_{N+1}^{({\rm fs})}(\beta;\xi)$ oscillates around a constant value and converges to $1/4$ when $N\rightarrow\infty$. Thus, the overall trend of $\Gamma_\xi^{(\mathrm{fs})}(N,d)$ exhibits a $(N+1)^{-3}$ dependence, as illustrated in Figs.~\ref{fig:fs_discrete}(d-f).  From these results, we can see that the analytical results are consistent with the exact numerical calculations very well, which clearly verify the validity of our analytical expression of $\Gamma_\xi^{(\mathrm{fs})}(N,d)$.

\subsubsection{Total linewidth of the most subradiant modes in the deep-subwavelength regime}

Combining the contributions from the guided-mode part Eq. \eqref{eq:Gamma-1D-final} and the free-space part Eq. \eqref{eq:Gamma-fs-analytic}, the total decay rates of the most subradiant modes in the limit $N\gg1$, $\xi\ll N$, and $k_0d\ll1$ take the form
\begin{equation}
    \begin{aligned}
        \Gamma_\xi(N,d)\;&\approx \;\frac{\pi^2\xi^2}{(N+1)^3}
\big\{\frac{\Gamma}{4}[1+(-1)^{N+\xi}\cos(\theta_{N+1})] \\
&+\frac{\gamma}{4}[1+(-1)^{N+\xi}\mathcal{K}_{\rm fs}(\theta_{N+1})]\big\}.
      \label{eq:Gamma-final}
    \end{aligned}
\end{equation}

Both contributions share the characteristic $(N+1)^{-3}$ envelope of open-boundary Bragg-edge subradiant modes,
while the prefactors and oscillatory dependence on $(N+1)k_0d$ encode, 
respectively, guided-mode interference and finite-length free-space leakage.

\begin{figure}[!htbp]
    \centering
    \includegraphics[width=0.9\linewidth]{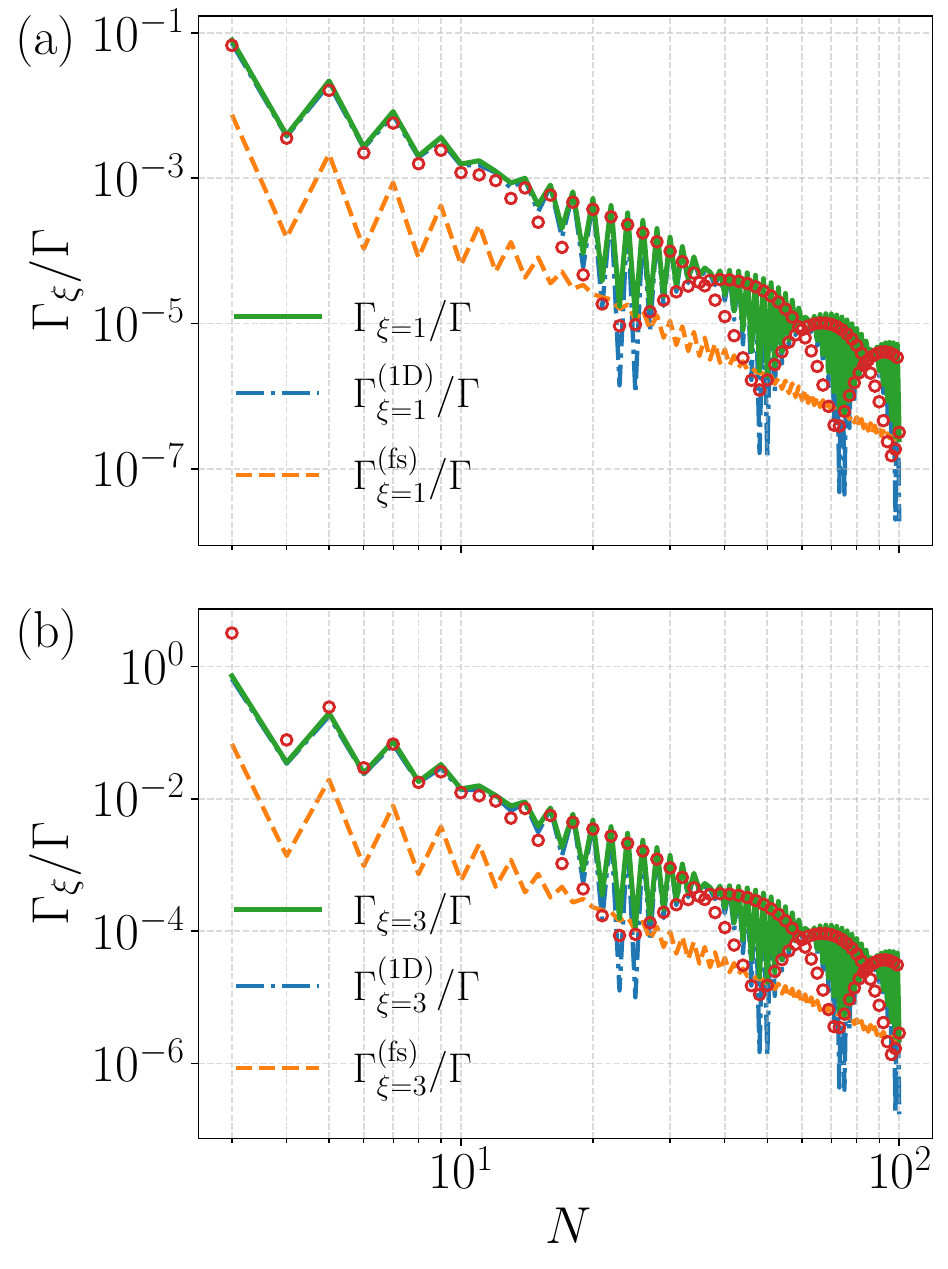}
    \caption{
    Analytical decomposition of the most subradiant decay rates in the nonideal-waveguide case with $d=0.02\lambda$ and $\gamma=0.1\Gamma$.
    (a) $\Gamma_\xi/\Gamma$ versus $N$ for $\xi=1$.
    (b) $\Gamma_\xi/\Gamma$ versus $N$ for $\xi=3$.
    In each panel, the dash-dotted blue line denotes the guided-mode contribution $\Gamma_\xi^{(\mathrm{1D})}/\Gamma$ from Eq.~\eqref{eq:Gamma-1D-final}, the dashed orange line denotes the free-space contribution $\Gamma_\xi^{(\mathrm{fs})}/\Gamma$ from Eq.~\eqref{eq:Gamma-fs-analytic}, and the  solid green line denotes the total analytical linewidth $\Gamma_\xi/\Gamma$ from Eq.~\eqref{eq:Gamma-final}.
    The red open circles are the exact numerical results.
    The figure shows that both the guided and nonguided channels inherit the same overall $(N+1)^{-3}$ subradiant envelope, while their different oscillatory structures combine to produce the full finite-size linewidth pattern.
    }
    \label{fig:linewidth_nonideal}
\end{figure}

As shown in Fig.~\ref{fig:linewidth_nonideal}, the analytical decomposition of the nonideal-waveguide linewidth agrees well with the numerical eigenvalues for both $\xi=1$ and $\xi=3$.
For each mode index $\xi$ within the Bragg-edge family, the full linewidth is reproduced by the sum of the guided contribution $\Gamma_\xi^{(\mathrm{1D})}$ and the free-space contribution $\Gamma_\xi^{(\mathrm{fs})}$.
This confirms that Eqs.~\eqref{eq:Gamma-1D-final}, \eqref{eq:Gamma-fs-analytic}, and \eqref{eq:Gamma-final} correctly capture not only the overall $(N+1)^{-3}$ subradiant envelope, but also the detailed finite-size oscillatory structure of the linewidth.

A clear physical picture also emerges from Fig.~\ref{fig:linewidth_nonideal}.
The guided-mode part exhibits the stronger parity-sensitive oscillations, whereas the free-space part provides a smoother positive background with the same overall $(N+1)^{-3}$ scaling.
Their sum therefore retains the Bragg-edge suppression while displaying a pronounced even--odd modulation.
The same mechanism is visible for both $\xi=1$ and $\xi=3$, with the higher-index mode showing a larger overall scale, consistent with the expected $\xi$ dependence within the low-lying Bragg-edge family.
The even--odd oscillations are a coherent finite-size interference effect and are therefore sensitive to phase fluctuations caused by positional disorder or atomic motion. 
A detailed analysis of the robustness of this effect, including numerical simulations for longitudinal position disorder and its connection to the Debye--Waller factor and the Lamb-Dicke regime, is presented in Sec.~\ref{sec:positional-disorder}.

\subsection{Real part: collective energy shift of the most subradiant modes}
\label{sec:J-derivation}

We now derive the real part $J_\xi$ of the collective eigenvalue within the
same Bragg-edge framework used above for the linewidth.
In contrast to $\Gamma_\xi$, which is suppressed as $(N+1)^{-3}$ by destructive
interference, the collective energy shift is dominated 
by the near-field free-space interaction in the deep-subwavelength regime and therefore approaches a finite
asymptotic value as $N\to\infty$, with only a weak $(N+1)^{-2}$ correction for
the Bragg-edge modes.

As shown in Eq.~\eqref{eq:Jtotal_main}, the collective energy shift $J_\xi$ can also be decomposed as two parts $J_\xi^{(\mathrm{1D})}$ and $J_\xi^{(\mathrm{fs})}$. In the following, we calculate $J_\xi^{(\mathrm{1D})}$ and $J_\xi^{(\mathrm{fs})}$ separately.

\subsubsection{Calculation of $J_\xi^{(\mathrm{1D})}$}

The energy shift due to the waveguide modes is similar to the left-branch ideal-waveguide result in Eq.~\eqref{eq:J_ideal_main_equiv}, which is the relevant branch in the deep-subwavelength regime considered below.
For the finite open chain, if the open-boundary Dirichlet quantization is adopted, the Bragg-edge variable is $a=\pi\xi/(N+1)$,
so that the finite-size formulas are obtained by the replacement $N\to N+1$, and the energy shift is given by
\begin{equation}
  J_\xi^{(\mathrm{1D})}(N,d)
  \approx
  -\frac{\Gamma}{2}\tan\!\left(\frac{k_0d}{2}\right)
  -\frac{\Gamma}{8}
  \frac{\sin(k_0d/2)}{\cos^3(k_0d/2)}
  \left(\frac{\pi\xi}{N+1}\right)^2.
  \label{eq:J_nonideal_main_equiv}
\end{equation}
When $N\rightarrow\infty$, $J_\xi^{(\mathrm{1D})}(N,d)\rightarrow -(\Gamma/2)\tan\!\left(k_0d/2\right)$. Apart from this $N$-independent background, the leading finite-size correction to the guided-mode energy shift is proportional to $\xi^2/(N+1)^2$. This should be contrasted with the linewidth, whose leading subradiant scaling is proportional to $\xi^2/(N+1)^3$.

In the deep-subwavelength regime $k_0d\ll1$, it can be further reduced to
\begin{equation}
  J_\xi^{(\mathrm{1D})}(N,d)
  \approx
  -\frac{\Gamma}{4}k_0d
  -\frac{\Gamma}{16}
  \left(\frac{\pi\xi}{N+1}\right)^2 k_0d
  +\cdots,
  \label{eq:J1D-smallbeta-main}
\end{equation}
When $N\rightarrow\infty$, $J_\xi^{(\mathrm{1D})}(N,d)\rightarrow -\Gamma k_0d/4$, which vanishes when $d\rightarrow 0$.

\subsubsection{Calculation of $J_\xi^{(\mathrm{fs})}$}

The collective energy shift due to the free-space modes is given by Eq. \eqref{eq:Jfs_main}. 
Since the kernel \(\mathcal{L}_{\mathrm{fs}}\!\big(k_0 z_{jl}\big)\) depends only on the separation \(z_{jl}=|j-l|d\), we may use the same discrete autocorrelation function as in the linewidth problem,
\begin{equation} 
\mathcal{C}_\xi(\Delta) = \sum_{n=1}^{N-\Delta} c_\xi^{*}(n+\Delta)c_\xi(n). \label{eq:autocorrelation-J} 
\end{equation}
For the Bragg-edge Dirichlet modes considered here, \(\mathcal C_\xi(\Delta)\) is real. Therefore, the free-space energy shift in Eq.~\eqref{eq:Jfs_main} reduces to the single sum
\begin{equation}
  J_\xi^{(\mathrm{fs})}(N,d)
  =
  \gamma\sum_{\Delta=1}^{N-1}
  \mathcal{L}_{\mathrm{fs}}(\beta\Delta)\,
  \mathcal{C}_\xi(\Delta),
  \qquad
  \beta\equiv k_0 d.
  \label{eq:Jfs-single-main}
\end{equation}

For the Bragg-edge modes,  $a\ll 1$ and in the deep-subwavelength regime $\beta\ll1$. Under these conditions, we can obtain (see Appendix~\ref{app:Jfs} for a detailed derivation)
\begin{equation}
  \begin{aligned}
  J_\xi^{(\mathrm{fs})}(N,d)
  \approx\;&
  -\frac{9\gamma}{8}\frac{\zeta(3)}{(k_0d)^3}
  +\frac{3\gamma\ln2}{4}\frac{1}{k_0d} \\
  &+\frac{3\gamma\pi^2\xi^2\ln2}{4}
  \frac{1}{(N+1)^2}\frac{1}{(k_0d)^3} \\ 
  &
  +O\!\left(\gamma k_0d\right)
  +O\!\left(\gamma\frac{\xi^2}{(N+1)^2}\frac{1}{k_0d}\right).
  \end{aligned}
  \label{eq:Jfs-final-main}
\end{equation}
The first term is the dominant near-field contribution, scaling as $d^{-3}$.
Unlike the linewidth, this leading term does not vanish as $N$ increases.
Instead, for the Bragg-edge modes the first nontrivial finite-size effect
enters at order $\xi^2/(N+1)^2$.

It is convenient to define the asymptotic free-space energy shift ($N\rightarrow\infty$)
\begin{equation}
  J_\infty^{(\mathrm{fs})}(d)
  \approx
  -\frac{9\gamma}{8}\frac{\zeta(3)}{(k_0d)^3}
  +\frac{3\gamma\ln2}{4}\frac{1}{k_0d}
  +O(\gamma k_0d),
  \label{eq:Jfs-infty-main}
\end{equation}
so that, to the leading $\xi$-dependent order retained here,
\begin{equation}
  J_\xi^{(\mathrm{fs})}(N,d)
  \approx
  J_\infty^{(\mathrm{fs})}(d)
  +\frac{3\gamma\pi^2\xi^2\ln2}{4}
  \frac{1}{(N+1)^2}\frac{1}{(k_0d)^3}.
  \label{eq:Jfs-finite-main}
\end{equation}

\begin{figure*}[!htbp]
    \centering
    \includegraphics[width=0.98\linewidth]{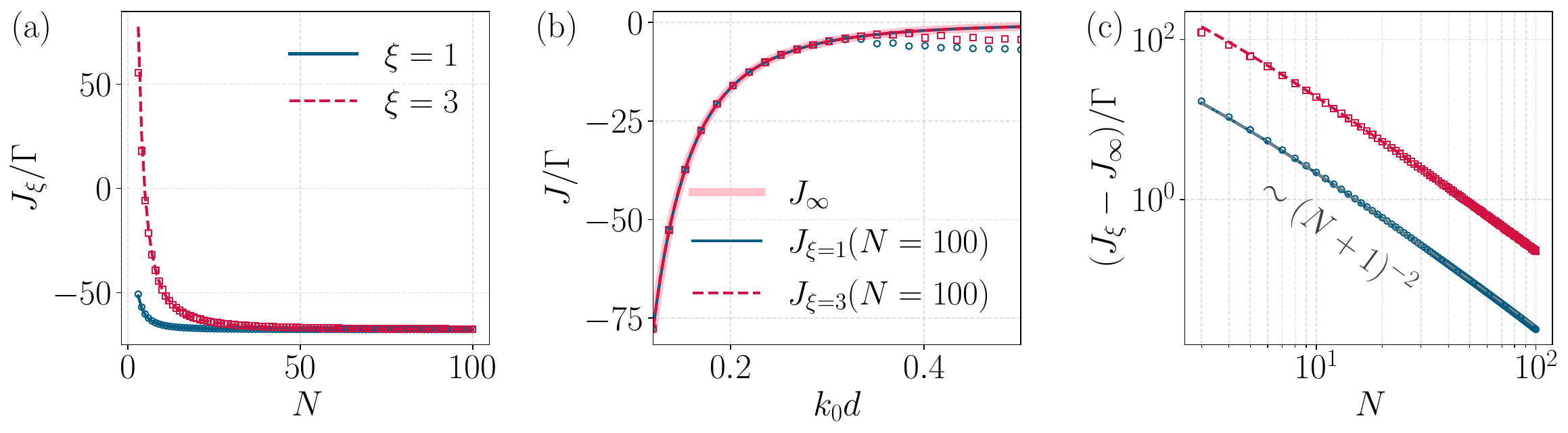}
    \caption{
Comparison between the numerical eigenvalues of the effective Hamiltonian and the analytical asymptotic formulas for the collective energy shift in the nonideal-waveguide case with $\gamma=0.1\Gamma$.
(a) collective energy shifts $J_\xi/\Gamma$ versus $N$ at fixed $d=0.02\lambda$ for $\xi=1$ (solid blue line and open circles) and $\xi=3$ (dashed red line and open squares).
(b) $J/\Gamma$ versus $k_0d$ at fixed $N=100$, showing the thermodynamic-limit shift $J_\infty$ from Eq.~\eqref{eq:Jinf-main} (thick pink line), together with the finite-$N$ results for $\xi=1$ (solid blue line and open circles) and $\xi=3$ (dashed red line and open squares).
(c) Deviation $(J_\xi-J_\infty)/\Gamma$ versus $N$ at fixed $d=0.02\lambda$ on a log--log scale for $\xi=1$ and $\xi=3$; the gray guide indicates the expected $(N+1)^{-2}$ scaling from Eq.~\eqref{eq:Jxi-Jinf}.
In all panels, lines denote analytical predictions and symbols denote numerical eigenvalues.
}
    \label{fig:shift_nonideal}
\end{figure*}

\subsubsection{Final asymptotic form}

Combining Eqs.~\eqref{eq:J_nonideal_main_equiv} and \eqref{eq:Jfs-finite-main}, we
obtain the asymptotic collective energy shift of the most subradiant modes,
\begin{equation}
  J_\xi(N,d)
  \approx
  J_\infty(d)
  +
  \frac{\pi^2\xi^2 }{(N+1)^2}C(d)
  +\cdots,
  \label{eq:Jtotal-asym-main}
\end{equation}
with
\begin{equation}
  J_\infty(d)
  \approx
  -\frac{9\gamma}{8}\frac{\zeta(3)}{(k_0d)^3}
  +\frac{3\gamma\ln2}{4}\frac{1}{k_0d}
  -\frac{\Gamma}{2}\tan\!\left(\frac{k_0d}{2}\right).
  \label{eq:Jinf-main}
\end{equation}
and
\begin{equation}
  C(d)   \approx
    \frac{3\gamma\ln2}{4(k_0d)^3}
    -\frac{\Gamma}{8}
    \frac{\sin(k_0d/2)}{\cos^3(k_0d/2)}.
  \label{eq:C-main}
\end{equation}

Equivalently,
\begin{equation}
  J_\xi(N,d)-J_\infty(d)
  \propto
  \frac{\xi^2}{(N+1)^2}
  \label{eq:Jxi-Jinf}
\end{equation}
for the Bragg-edge modes.

Equation~\eqref{eq:Jtotal-asym-main} summarizes the central contrast between the
real and imaginary parts of the subradiant eigenvalue.
While the linewidth narrows as $(N+1)^{-3}$ due to Bragg-edge destructive
interference, the collective energy shift is dominated by the free-space near field and
therefore approaches a finite asymptotic value as $N$ increases, with only a
weak $\xi^2/(N+1)^2$ correction.

\subsubsection{Numerical verification}

The behavior of the collective energy shift is summarized in Fig.~\ref{fig:shift_nonideal}.
At fixed deep-subwavelength spacing, the two low-lying Bragg-edge modes with $\xi=1$ and $\xi=3$ approach the same asymptotic value as $N$ increases, while the $\xi=3$ mode shows a visibly larger finite-size offset as is shown in Fig.~\ref{fig:shift_nonideal}(a).
This is precisely the behavior predicted by Eq.~\eqref{eq:Jtotal-asym-main}: the dominant contribution is $\xi$ independent and survives in the thermodynamic limit, while the mode-index dependence enters only through the correction proportional to $\xi^2/(N+1)^2$.

Figure~\ref{fig:shift_nonideal}(b) further shows that the spacing dependence of the converging collective energy shift is captured to high accuracy by the thermodynamic-limit expression $J_\infty$ in Eq.~\eqref{eq:Jinf-main} especially when $k_0d$ is not very large.
The finite-$N$ curves for $\xi=1$ and $\xi=3$ remain close to $J_\infty$ over the whole plotted interval, indicating that the dominant shift is set primarily by the near-field free-space interaction rather than by branch-dependent finite-size effects.
In Fig.~\ref{fig:shift_nonideal}(c), the finite-size dependence is isolated by plotting $(J_\xi-J_\infty)/\Gamma$ on a log--log scale.
The numerical data for both $\xi=1$ and $\xi=3$ fall on straight lines with slope $-2$, in agreement with the predicted $(N+1)^{-2}$ scaling, while the $\xi=3$ branch lies systematically above the $\xi=1$ branch because of the $\xi^2$ dependence of the prefactor.

\section{Influence of positional disorder}
\label{sec:positional-disorder}

\begin{figure}[!htbp]
    \centering
    \includegraphics[width=0.98\linewidth]{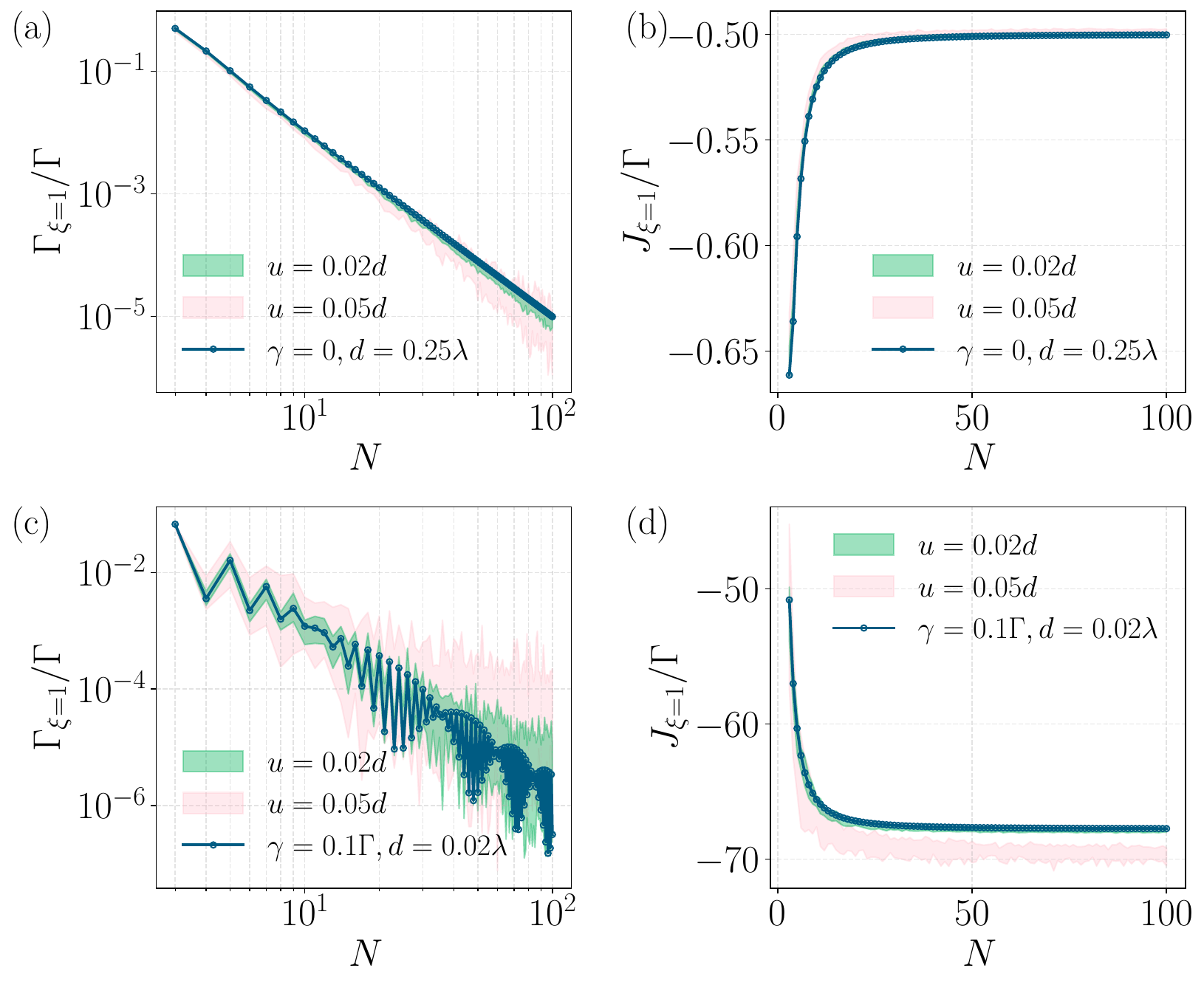}
    \caption{ Robustness of the most subradiant eigenvalue against longitudinal positional disorder.
  The emitter positions are modeled as \(z_j=(j-1)d+u\epsilon_j\), where \(\epsilon_j\) is uniformly distributed in \([-1,1]\).The shaded regions indicate the minimum-to-maximum range over 20 disorder realizations for \(u=0.02d\) and \(u=0.05d\).
  (a) Linewidth \(\Gamma_{\xi=1}/\Gamma\) versus \(N\) for the ideal-waveguide case with \(d=0.25\lambda\) and \(\gamma=0\).
  (b) Collective energy shift \(J_{\xi=1}/\Gamma\) for the same ideal-waveguide case.
  (c) Linewidth \(\Gamma_{\xi=1}/\Gamma\) versus \(N\) for the nonideal deep-subwavelength case with \(d=0.02\lambda\) and \(\gamma=0.1\Gamma\).
  (d) Collective energy shift \(J_{\xi=1}/\Gamma\) for the same nonideal case.
  }
    \label{fig:disorder-robustness}
\end{figure}

We now investigate the robustness of the behavior of collective subradiant modes against longitudinal positional disorder. Such disorder can arise from finite trapping temperature, imperfect emitter localization, or fabrication inaccuracies in solid-state implementations. 
To make this point quantitative, we perform numerical simulations to discuss the influence of position disorder. Specifically, we model the position of the \(j\)th emitter as
\begin{equation}
  z_j=(j-1)d+u\epsilon_j ,
  \label{eq:disorder-model}
\end{equation}
where \(\epsilon_j\) is an independent random variable uniformly distributed in \([-1,1]\), and \(u\) is the maximum disorder amplitude. For each atom number \(N\), we performed 20 disorder realizations for two representative disorder strengths ( \(u=0.02d\) and \(u=0.05d\)), and the results are shown in Fig.~\ref{fig:disorder-robustness} where both the ideal and nonideal waveguides are considered. 

The numerical results demonstrate that the ideal-waveguide subradiant branch is highly robust against both disorder strengths. The linewidth preserves the overall $N^{-3}$ subradiant scaling trend [Fig.~\ref{fig:disorder-robustness}(a)], and the collective energy shift remains close to the clean-chain result [Fig.~\ref{fig:disorder-robustness}(b)].

In the nonideal deep-subwavelength regime, the linewidth exhibits broader fluctuations under longitudinal positional disorder, as shown in Fig.~\ref{fig:disorder-robustness}(c). For the weaker disorder \(u=0.02d\), the even--odd oscillatory pattern remains clearly identifiable. For the larger disorder \(u=0.05d\), the shaded region becomes broader and partially masks the parity contrast, although the subradiant suppression and the overall decreasing trend with \(N\) are still preserved. This behavior reflects the coherent nature of the parity oscillation. For longitudinal fluctuations \(z_j=(j-1)d+\delta z_j\), the boundary-interference term acquires a random optical phase of order \(k_0(\delta z_N-\delta z_1)\), so that its contrast is controlled by the phase average \(\langle e^{ik_0(\delta z_N-\delta z_1)}\rangle\). For thermal Gaussian fluctuations with rms displacement \(\sigma_z\), this average gives the usual Debye--Waller-type factor \(\exp(-k_0^2\sigma_z^2)\)~\cite{Debye1913,Waller1923}. For the uniformly distributed disorder used here, \(\delta z_j=u\epsilon_j\) with \(\epsilon_j\in[-1,1]\), the rms displacement is \(\sigma_z=u/\sqrt{3}\), and the same weak-disorder contrast-reduction picture applies. 
Thus the parity oscillations are expected to remain visible in the Lamb-Dicke regime \(\eta=k_0\sigma_z\ll1\)~\cite{Dicke1953,Leibfried2003}. 
For \(d=0.02\lambda\), one has \(k_0u\simeq2.5\times10^{-3}\) for \(u=0.02d\), and \(k_0u\simeq6.3\times10^{-3}\) for \(u=0.05d\). 
Therefore, the rms Lamb-Dicke parameter \(\eta=k_0u/\sqrt{3}\) is very small, indicating that the optical phase coherence is not destroyed. 
 The stronger fluctuation in Fig.~\ref{fig:disorder-robustness}(c) is instead associated with the Bragg-edge structure of the subradiant mode. Since \(k_b\simeq\pi/d\), the same positional disorder produces a Bragg-phase fluctuation \(k_bu\simeq\pi u/d\), which is \(0.063\) for \(u=0.02d\) and \(0.157\) for \(u=0.05d\). Such relative lattice disorder can perturb the Bragg-edge standing-wave structure and mix nearby narrow subradiant modes. Because the clean linewidth is already extremely small, this produces a large fractional spread in \(\Gamma_{\xi=1}\), even though \(k_0u\ll1\). Therefore, position disorder gradually reduces the parity contrast, while the overall subradiant trend remains robust.

The collective energy shift in the nonideal case is more sensitive to disorder, as shown in Fig.~\ref{fig:disorder-robustness}(d). This sensitivity has a different physical origin from the linewidth fluctuations. As shown in Eq.~\eqref{eq:Jinf-main}, the asymptotic collective energy shift \(J_\infty(d)\) is dominated in the deep-subwavelength regime by the near-field free-space dipole--dipole interaction, whose leading dependence is negative and scales as \(J_\infty\sim -d^{-3}\). In a disordered array, the fixed nearest-neighbor separation \(d\) is effectively replaced by a fluctuating distance \(d+\delta r\). Although \(\delta r\) has zero mean, the inverse-cubic interaction does not average back to the clean value, because $\left\langle (d+\delta r)^{-3}\right\rangle > d^{-3}$.
Thus, configurations with slightly reduced separations enhance the negative near-field shift more strongly than configurations with increased separations suppress it. Since the leading near-field contribution to \(J_\infty\) carries a negative sign, the disorder-averaged energy shift is pushed toward more negative values. This explains why, for the larger disorder amplitude \(u=0.05d\), the shaded region in Fig.~\ref{fig:disorder-robustness}(d) is not only broadened but also clearly displaced downward. This behavior further illustrates the main distinction emphasized in this work: the linewidth is governed primarily by Bragg-edge destructive interference and phase coherence, whereas the collective energy shift is governed mainly by local near-field dipole--dipole interactions and is therefore more sensitive to short-distance positional fluctuations.

Experimentally, the most direct signature of the parity-dependent linewidth oscillation would be an alternating linewidth of the narrowest subradiant resonance when the atom number is changed from \(N\) to \(N+1\), or an oscillatory modulation of the linewidth when the product \((N+1)k_0d\) is tuned. In reflection or transmission measurements, this would appear as an even--odd modulation of the width and depth of the narrow subradiant resonance. The results in Fig.~\ref{fig:disorder-robustness} show that these signatures should remain observable for weak positional disorder in the Lamb-Dicke regime.

\section{Summary and discussion}
\label{sec:conclusion}

In summary, we have developed an analytical theory for the most subradiant collective modes in a finite one-dimensional emitter array coupled to a nonideal waveguide, where emitters interact simultaneously with a guided channel and with free-space radiation modes. By starting from the effective non-Hermitian Hamiltonian and employing a Bragg-edge open-boundary ansatz, we derived compact expressions for the full complex eigenvalue of these low-energy subradiant branches, thereby extending the standard theory of waveguide subradiance from a primary focus on the decay rate to a unified description encompassing both the linewidth and the collective energy shift. Our analysis reveals a sharp qualitative distinction between these two quantities: while the decay rate retains the characteristic $N^{-3}$ scaling even in the presence of nonguided couplings, it acquires a pronounced even--odd oscillatory structure in the deep-subwavelength regime arising from boundary interference, a feature we confirmed through both analytical derivation and exact numerical verification. In contrast, the collective energy shift is dominated by free-space near-field interactions, with its leading asymptotic behavior scaling as $-(k_0 d)^{-3}$ and the first finite-size correction scaling as $N^{-2}$, a parametric dependence that differs fundamentally from that of the linewidth.

Our calculations here have direct implications for waveguide-QED-based platforms in quantum information and metrology. The linewidth suppression directly quantifies the radiative protection efficiency, which is essential for extending the coherence time of on-chip atomic ensembles used as quantum memory elements and high-sensitive sensing. In parallel, the collective energy shift determines the spectral addressing fidelity and governs the phase-matching conditions for multi-qubit gates and long-range coherent interactions in spin models. Additionally, the even-odd oscillation of the linewidth as a function of atom number offers a non-destructive probe for detecting minute positional disorders—an important practical capability for in situ calibration of emitter arrays and for validating fabrication precision in nanophotonic devices.

The present derivation relies on a one-dimensional open-chain
Bragg-edge ansatz and on the associated one-dimensional structure factor.
This specific single-index sine ansatz cannot be transferred unchanged to
a higher-dimensional array.
Nevertheless, the central idea of our approach---constructing subradiant
branches from open-boundary modes near a radiative band edge and
evaluating the corresponding structure factor---can, in principle, be
extended to two-dimensional or higher-dimensional emitter arrays coupled
to waveguides or cavities.
For example, for a two-dimensional emitter array coupled to a network of
one-dimensional waveguides running along different directions, one may
construct higher-dimensional open-boundary modes as products of sine-like
Bragg-edge modes along the corresponding waveguide directions.
In such extensions, the one-dimensional autocorrelation sums should be
replaced by multidirectional structure factors and multidimensional
lattice Green-function sums for the specific geometry under
consideration.
The resulting theory would have to be adjusted to the additional spatial
dimensions and to the corresponding photonic environment, and may lead to
new and interesting finite-size effects.
Therefore, our method provides a starting point for higher-dimensional or
cavity-assisted generalizations, but the \(N^{-3}\) linewidth law and the
\(N^{-2}\) energy-shift correction derived here should be regarded as
specific consequences of the finite one-dimensional waveguide geometry
considered in this work.

\section{\label{sec:acknowledgments}Acknowledgments}
This work was supported by the National Key R\&D Program of China (Grant No. 2021YFA1400800), the Key Program of National Natural Science Foundation of China (Grant No. 12334017), Guangdong Provincial Quantum Science Strategic Initiative (Grant No. GDZX2505001 and GDZX2406001), and Guangdong Basic and Applied Basic Research Foundation (Grant No. 2026A1515011705).

\appendix

\section{Ideal waveguide: derivation of the linewidth and frequency shift}
\label{app:ideal-waveguide-derivation}

In this section we derive the two asymptotic subradiant families of an ideal finite waveguide and show explicitly how selecting the smaller linewidth produces the piecewise results in Eqs.~\eqref{eq:Gamma_zhang_main} and \eqref{eq:J_ideal_main_equiv}. We restrict the derivation to $0<d<\lambda/2$; the spectrum repeats with period $\lambda/2$.

\subsection{Effective Hamiltonian and Bloch-state action}

For an ideal waveguide, the decay into nonguided modes is neglected, so that $\gamma=0$. The effective non-Hermitian Hamiltonian in the single-excitation manifold is
\begin{equation}
H_{\mathrm{eff}}^{(1\mathrm{D})}
=
-\frac{i\Gamma}{2}
\sum_{m,n=1}^{N}
e^{ik_0|z_m-z_n|}
\sigma_m^\dagger \sigma_n ,
\label{eq:Heff_ideal_start}
\end{equation}
where $\sigma_m^\dagger = |e_m\rangle\langle g_m|$, and $\Gamma$ is the decay rate into the guided mode.

We start from the Bloch state
\begin{equation}
|k\rangle
=
\frac{1}{\sqrt{N}}
\sum_{\ell=1}^{N}
e^{ik z_\ell}|e_\ell\rangle .
\label{eq:Bloch_state_def}
\end{equation}
Acting with Eq.~\eqref{eq:Heff_ideal_start} on $|k\rangle$ gives
\begin{equation}
\begin{aligned}
H_{\mathrm{eff}}^{(1\mathrm{D})}|k\rangle
&=
-\frac{i\Gamma}{2}
\frac{1}{\sqrt{N}}
\sum_{m,n,\ell}
e^{ik_0|z_m-z_n|}
e^{ik z_\ell}
\sigma_m^\dagger \sigma_n |e_\ell\rangle \\
&=
-\frac{i\Gamma}{2}
\frac{1}{\sqrt{N}}
\sum_{m,n}
e^{ik_0|z_m-z_n|}
e^{ik z_n}
|e_m\rangle ,
\end{aligned}
\label{eq:Heff_on_k_step1}
\end{equation}
where we used
\begin{equation}
\sigma_m^\dagger \sigma_n |e_\ell\rangle
=
\delta_{n\ell}|e_m\rangle .
\end{equation}

For fixed $m$, split the sum over $n$ into the three regions $n<m$, $n=m$, and $n>m$. One finds
\begin{align}
&\sum_{n=1}^{N}
e^{ik_0|z_m-z_n|}
e^{ik z_n}  \nonumber \\
&=
e^{ik z_m}
\left[
1
+
\sum_{s=1}^{m-1} e^{i(k_0-k)sd}
+
\sum_{s=1}^{N-m} e^{i(k_0+k)sd}
\right].
\label{eq:sum_fixed_m}
\end{align}
Using the geometric-series identity
\begin{equation}
\sum_{s=1}^{L} e^{i\theta s}
=
\frac{e^{i\theta}\left(1-e^{i\theta L}\right)}{1-e^{i\theta}},
\label{eq:geom_series}
\end{equation}
and rearranging the result, one obtains the exact decomposition
\begin{equation}
H_{\mathrm{eff}}^{(1\mathrm{D})}|k\rangle
=
\omega_k |k\rangle
-\frac{i\Gamma}{2}
\left(
g_k |k_0\rangle
-
h_k |-k_0\rangle
\right),
\label{eq:Heff_Bloch_exact}
\end{equation}
where
\begin{equation}
\omega_k
=
\frac{\Gamma}{4}
\left[
\cot\!\left(\frac{(k_0+k)d}{2}\right)
+
\cot\!\left(\frac{(k_0-k)d}{2}\right)
\right],
\label{eq:omega_k_cot}
\end{equation}
and the tail coefficients are
\begin{equation}
g_k
=
\frac{e^{i(k-k_0)z_1}}{1-e^{i(k-k_0)d}},
\qquad
h_k
=
\frac{e^{i(k+k_0)z_N}}{e^{-i(k+k_0)d}-1}.
\label{eq:gk_hk_def}
\end{equation}
Equation~\eqref{eq:Heff_Bloch_exact} shows that a single Bloch state is not, in general, an eigenstate of $H_{\mathrm{eff}}^{(1\mathrm{D})}$ because of the residual coupling to the two superradiant states $|\pm k_0\rangle$.

\subsection{Tail cancellation and the two subradiant branches}

To eliminate the tails, consider the superposition
\begin{equation}
|\phi_k\rangle = A|k\rangle + B|-k\rangle .
\label{eq:phi_k_superposition}
\end{equation}
Using Eq.~\eqref{eq:Heff_Bloch_exact}, we obtain
\begin{equation}
\begin{aligned}
H_{\mathrm{eff}}^{(1\mathrm{D})}|\phi_k\rangle
&=
\omega_k |\phi_k\rangle \\
-\frac{i\Gamma}{2} 
&
\Big[
(A g_k + B g_{-k})|k_0\rangle
-
(A h_k + B h_{-k})|-k_0\rangle
\Big].
\end{aligned}
\end{equation}
Thus the tail-free condition requires
\begin{equation}
A g_k + B g_{-k} = 0,
\qquad
A h_k + B h_{-k} = 0.
\end{equation}
A nontrivial solution exists only if
\begin{equation}
g_k h_{-k} = g_{-k} h_k .
\label{eq:tail_cancellation_condition}
\end{equation}

Equation~\eqref{eq:tail_cancellation_condition} has two families of
subradiant solutions, one near the Brillouin-zone center and the other
near its edge. Expanding the tail-cancellation condition to the first
nonvanishing imaginary order gives~\cite{zhang2019a}
\begin{equation}
\begin{aligned}
k_{\xi}^{(0)}d
&\approx
\frac{\pi\xi}{N}
\left[
1-\frac{i}{N}
\cot\!\left(\frac{k_0d}{2}\right)
\right],
&& k_{\xi}^{(0)}\approx0,
\\
\left(k_{\xi}^{(\pi)}+\frac{\pi}{d}\right)d
&\approx
\frac{\pi\xi}{N}
\left[
1+\frac{i}{N}
\tan\!\left(\frac{k_0d}{2}\right)
\right],
&& k_{\xi}^{(\pi)}\approx-\frac{\pi}{d},
\end{aligned}
\label{eq:ideal_two_k_branches_app}
\end{equation}
where $\xi=1,2,\ldots$ and $\xi\ll N$. The real parts provide the
finite-size quantization at order $N^{-1}$, whereas the imaginary
corrections are of order $N^{-2}$ and generate linewidths of order
$N^{-3}$.

\subsection{Closed-form dispersion and the two asymptotic families}

Using
\begin{equation}
\cot A+\cot B
=
\frac{\sin(A+B)}{\sin A\sin B},
\end{equation}
the exact dispersion in Eq.~\eqref{eq:omega_k_cot} can be written as
\begin{equation}
\omega_k
=
\frac{\Gamma}{2}
\frac{\sin(k_0d)}
{\cos(kd)-\cos(k_0d)}.
\label{eq:omega_k_closed}
\end{equation}

\subsubsection{Family near the Brillouin-zone center}

For $k_{\xi}^{(0)}\approx0$,
\begin{equation}
\cos\!\left(k_{\xi}^{(0)}d\right)
\approx
1-\frac{\left(k_{\xi}^{(0)}d\right)^2}{2}.
\end{equation}
Substitution into Eq.~\eqref{eq:omega_k_closed} gives
\begin{equation}
\omega_\xi^{(0)}
\approx
\frac{\Gamma}{2}
\cot\!\left(\frac{k_0d}{2}\right)
+
\frac{\Gamma}{8}
\frac{\cos(k_0d/2)}
{\sin^3(k_0d/2)}
\left(k_{\xi}^{(0)}d\right)^2.
\label{eq:omega_center_before_substitution}
\end{equation}
From Eq.~\eqref{eq:ideal_two_k_branches_app},
\begin{equation}
\left(k_{\xi}^{(0)}d\right)^2
\approx
\left(\frac{\pi\xi}{N}\right)^2
\left[
1-\frac{2i}{N}
\cot\!\left(\frac{k_0d}{2}\right)
\right].
\end{equation}
Consequently,
\begin{align}
\omega_\xi^{(0)}
\approx{}&
\frac{\Gamma}{2}
\cot\!\left(\frac{k_0d}{2}\right)
+
\frac{\Gamma}{8}
\left(\frac{\pi\xi}{N}\right)^2
\frac{\cos(k_0d/2)}
{\sin^3(k_0d/2)}
\nonumber\\
&{}
-i\frac{\Gamma}{4}
\frac{\pi^2\xi^2}{N^3}
\frac{\cos^2(k_0d/2)}
{\sin^4(k_0d/2)}.
\label{eq:omega_center_final}
\end{align}
Writing
\begin{equation}
\omega_\xi^{(0)}
=
J_\xi^{(0)}-\frac{i}{2}\Gamma_\xi^{(0)},
\end{equation}
we identify
\begin{equation}
J_\xi^{(0)}
\approx
\frac{\Gamma}{2}
\cot\!\left(\frac{k_0d}{2}\right)
+
\frac{\Gamma}{8}
\left(\frac{\pi\xi}{N}\right)^2
\frac{\cos(k_0d/2)}
{\sin^3(k_0d/2)}
\label{eq:J_center_app}
\end{equation}
and
\begin{equation}
\Gamma_\xi^{(0)}
\approx
\frac{\Gamma}{2}
\frac{\pi^2\xi^2}{N^3}
\frac{\cos^2(k_0d/2)}
{\sin^4(k_0d/2)}.
\label{eq:Gamma_center_app}
\end{equation}

\subsubsection{Family near the Brillouin-zone edge}

For $k_{\xi}^{(\pi)}\approx-\pi/d$,
\begin{equation}
\cos\!\left(k_{\xi}^{(\pi)}d\right)
\approx
-1+
\frac{
\left[
\left(k_{\xi}^{(\pi)}+\pi/d\right)d
\right]^2
}{2}.
\end{equation}
Substitution into Eq.~\eqref{eq:omega_k_closed} gives
\begin{equation}
\omega_\xi^{(\pi)}
\approx
-\frac{\Gamma}{2}
\tan\!\left(\frac{k_0d}{2}\right)
-\frac{\Gamma}{8}
\frac{\sin(k_0d/2)}
{\cos^3(k_0d/2)}
\left[
\left(k_{\xi}^{(\pi)}+\frac{\pi}{d}\right)d
\right]^2.
\label{eq:omega_edge_before_substitution}
\end{equation}
Equation~\eqref{eq:ideal_two_k_branches_app} yields
\begin{equation}
\left[
\left(k_{\xi}^{(\pi)}+\frac{\pi}{d}\right)d
\right]^2
\approx
\left(\frac{\pi\xi}{N}\right)^2
\left[
1+\frac{2i}{N}
\tan\!\left(\frac{k_0d}{2}\right)
\right].
\end{equation}
It follows that
\begin{align}
\omega_\xi^{(\pi)}
\approx{}&
-\frac{\Gamma}{2}
\tan\!\left(\frac{k_0d}{2}\right)
-\frac{\Gamma}{8}
\left(\frac{\pi\xi}{N}\right)^2
\frac{\sin(k_0d/2)}
{\cos^3(k_0d/2)}
\nonumber\\
&{}
-i\frac{\Gamma}{4}
\frac{\pi^2\xi^2}{N^3}
\frac{\sin^2(k_0d/2)}
{\cos^4(k_0d/2)}.
\label{eq:omega_edge_final}
\end{align}
Hence,
\begin{equation}
J_\xi^{(\pi)}
\approx
-\frac{\Gamma}{2}
\tan\!\left(\frac{k_0d}{2}\right)
-\frac{\Gamma}{8}
\left(\frac{\pi\xi}{N}\right)^2
\frac{\sin(k_0d/2)}
{\cos^3(k_0d/2)}
\label{eq:J_edge_app}
\end{equation}
and
\begin{equation}
\Gamma_\xi^{(\pi)}
\approx
\frac{\Gamma}{2}
\frac{\pi^2\xi^2}{N^3}
\frac{\sin^2(k_0d/2)}
{\cos^4(k_0d/2)}.
\label{eq:Gamma_edge_app}
\end{equation}

\subsection{Selection of the narrowest family}

The ratio of the two linewidths is
\begin{equation}
\frac{\Gamma_\xi^{(\pi)}}
{\Gamma_\xi^{(0)}}
=
\tan^6\!\left(\frac{k_0d}{2}\right).
\label{eq:linewidth_branch_ratio}
\end{equation}
Within $0<d<\lambda/2$, this ratio is smaller than unity for
$0<d<\lambda/4$ and larger than unity for
$\lambda/4<d<\lambda/2$. Therefore, the linewidth-selected result is
\begin{equation}
\Gamma_\xi(N,d)
\approx
\frac{\Gamma}{2}
\frac{\pi^2\xi^2}{N^3}
\begin{cases}
\displaystyle
\frac{\sin^2(k_0d/2)}
{\cos^4(k_0d/2)},
&
0<d<\dfrac{\lambda}{4},
\\[1.0em]
\displaystyle
\frac{\cos^2(k_0d/2)}
{\sin^4(k_0d/2)},
&
\dfrac{\lambda}{4}<d<\dfrac{\lambda}{2},
\end{cases}
\label{eq:Gamma_ideal_Nplus1}
\end{equation}
and the corresponding collective shift is
\begin{equation}
J_\xi(N,d)
\approx
\begin{cases}
\begin{aligned}
&-\frac{\Gamma}{2}
\tan\!\left(\frac{k_0d}{2}\right)
\\[-0.2em]
&-\frac{\Gamma}{8}
\left(\frac{\pi\xi}{N}\right)^2
\frac{\sin(k_0d/2)}
{\cos^3(k_0d/2)}
\end{aligned}
&
0<d<\dfrac{\lambda}{4},
\\[1.4em]
\begin{aligned}
&\frac{\Gamma}{2}
\cot\!\left(\frac{k_0d}{2}\right)
\\[-0.2em]
&+\frac{\Gamma}{8}
\left(\frac{\pi\xi}{N}\right)^2
\frac{\cos(k_0d/2)}
{\sin^3(k_0d/2)}
\end{aligned}
&
\dfrac{\lambda}{4}<d<\dfrac{\lambda}{2}.
\end{cases}
\label{eq:J_ideal_equiv_Nplus1}
\end{equation}
At $d=\lambda/4$, the two linewidths are equal, whereas
$J_\xi^{(0)}=-J_\xi^{(\pi)}$. Consequently, the narrowest linewidth is
continuous through the crossing, but its associated collective shift
is not uniquely selected at that point.

\subsection{Deep-subwavelength limit}

In the deep-subwavelength regime $k_0d\ll1$,
\begin{equation}
\tan\!\left(\frac{k_0d}{2}\right)\approx \frac{k_0d}{2},
\qquad
\sec^2\!\left(\frac{k_0d}{2}\right)\approx 1.
\end{equation}
Then Eqs.~\eqref{eq:J_ideal_equiv_Nplus1} and \eqref{eq:Gamma_ideal_Nplus1} reduce to
\begin{equation}
J_\xi(N,d)
\approx
-\frac{\Gamma}{4}k_0d
-\frac{\Gamma}{16}
\left(\frac{\pi\xi}{N}\right)^2
k_0d
\label{eq:J_ideal_small_beta_Nplus1}
\end{equation}
and
\begin{equation}
\Gamma_\xi(N,d)
\approx
\frac{\Gamma}{8}
\frac{\pi^2\xi^2}{N^3}
(k_0d)^2
\label{eq:Gamma_ideal_small_beta_Nplus1}
\end{equation}
for $\xi\ll N$.

\section{Angular-average representation of $\mathcal{K}_{\mathrm{fs}}$}
\label{app:Kfs}

In this appendix, we verify the angular-average representation of $\mathcal{K}_{\mathrm{fs}}$ shown in Eq.~\eqref{eq:Kfs-angular}. 

Since $1+\mu^2$ is even in $\mu$, the imaginary part of the integrand in Eq.~\eqref{eq:Kfs-angular} (proportional to $\sin(x\mu)$) integrates to zero,
and we obtain
\begin{equation}
  \int_{-1}^{1}\!\mathrm{d}\mu\,(1+\mu^2)\,e^{i x\mu}
  = 2\int_{0}^{1}\!\mathrm{d}\mu\,(1+\mu^2)\cos(x\mu).
\end{equation}
Define
\begin{equation}
  I(x)
  \equiv \int_{0}^{1}\!\mathrm{d}\mu\,(1+\mu^2)\cos(x\mu)
  = I_0(x)+I_2(x),
\end{equation}
with
\begin{equation}
  I_0(x)=\int_{0}^{1}\cos(x\mu)\,\mathrm{d}\mu
  = \frac{\sin x}{x},
\end{equation}
and
\begin{equation}
  I_2(x)=\int_{0}^{1}\mu^2\cos(x\mu)\,\mathrm{d}\mu.
\end{equation}
Performing two integrations by parts for $I_2(x)$ yields
\begin{equation}
  I_2(x)
  = \frac{\sin x}{x} + \frac{2\cos x}{x^2} - \frac{2\sin x}{x^3}.
\end{equation}
Hence
\begin{equation}
  I(x)
  = I_0(x)+I_2(x)
  = \frac{2\sin x}{x} + \frac{2\cos x}{x^2} - \frac{2\sin x}{x^3}.
\end{equation}
Substituting this back into Eq.~\eqref{eq:Kfs-angular}, we find
\begin{equation}
\begin{aligned}
  \mathcal{K}_{\rm fs}(x)
  &= \frac{3}{8}\cdot 2I(x)
   = \frac{3}{4}\left[
        \frac{2\sin x}{x}
        +\frac{2\cos x}{x^2}
        -\frac{2\sin x}{x^3}
      \right]  \\
  &= \frac{3}{2}\left[
        \frac{\sin x}{x}
        +\frac{\cos x}{x^2}
        -\frac{\sin x}{x^3}
      \right],
\end{aligned}
\end{equation}
in complete agreement with Eq.~\eqref{eq:Kfs}. Thus the
angular-average representation \eqref{eq:Kfs-angular} is mathematically
equivalent to the original closed-form expression of the kernel.

\section{Angular integral for the free-space linewidth}
\label{app:angular_integral}

Here we evaluate the angular integral entering the free-space contribution to
the linewidth of the Bragg-edge subradiant mode.

Starting from Eq.~\eqref{eq:Gamma-fs-Sxi} and substituting the leading-order result Eq.~\eqref{eq:Sxi2-final}, 
we obtain
\begin{equation}
\Gamma_\xi^{(\mathrm{fs})}
\approx
\frac{\pi^2\xi^2}{(N+1)^3}\,
\frac{3\gamma}{16}\,
\mathcal{I}_\xi(\theta_{N+1}),
\label{eq:Gammafs_step}
\end{equation}
where
\begin{equation}
\mathcal{I}_\xi(\theta)
\equiv
\int_0^1 d\mu\,(1+\mu^2)
\Big[1+(-1)^{N+\xi}\cos(\theta\mu)\Big].
\label{eq:Itheta_def1}
\end{equation}
and
\begin{equation}
  \theta_{N+1}\equiv (N+1)k_0d .
\end{equation}

The two $\mu$ integrals are evaluated separately. 
The first integral is elementary:
\begin{equation}
  \int_0^1 d\mu\,(1+\mu^2)
  =
  \left[\mu+\frac{\mu^3}{3}\right]_0^1
  = \frac{4}{3}.
\end{equation}

For the second integral, define
\begin{equation}
  I_1(\theta)\equiv \int_0^1 d\mu\,(1+\mu^2)\cos(\theta\mu).
\end{equation}
It can be evaluated as
\begin{equation}
\begin{aligned}
  I_1(\theta)
  &= \int_0^1 d\mu\,\cos(\theta\mu)
   + \int_0^1 d\mu\,\mu^2\cos(\theta\mu).
\end{aligned}
\end{equation}
The first term is
\begin{equation}
  \int_0^1 d\mu\,\cos(\theta\mu)
  = \frac{\sin\theta}{\theta}.
\end{equation}
For the second term, integrating by parts twice gives
\begin{equation}
\begin{aligned}
  \int_0^1 d\mu\,\mu^2\cos(\theta\mu)
  &=
  \left.\mu^2\frac{\sin(\theta\mu)}{\theta}\right|_0^1
  -\frac{2}{\theta}\int_0^1 d\mu\,\mu\sin(\theta\mu) \\
  &=
  \frac{\sin\theta}{\theta}
  +\frac{2\cos\theta}{\theta^2}
  -\frac{2\sin\theta}{\theta^3}.
\end{aligned}
\end{equation}
Therefore,
\begin{equation}
\begin{aligned}
  I_1(\theta)
  &=
  \frac{\sin\theta}{\theta}
  +\left(
    \frac{\sin\theta}{\theta}
    +\frac{2\cos\theta}{\theta^2}
    -\frac{2\sin\theta}{\theta^3}
  \right) \\
  &=
  2\left[
    \frac{\sin\theta}{\theta}
    +\frac{\cos\theta}{\theta^2}
    -\frac{\sin\theta}{\theta^3}
  \right].
\end{aligned}
\end{equation}
Using the definition of the free-space kernel,
\begin{equation}
  \mathcal{K}_{\rm fs}(\theta)
  = \frac{3}{2}\left[
      \frac{\sin\theta}{\theta}
      +\frac{\cos\theta}{\theta^2}
      -\frac{\sin\theta}{\theta^3}
    \right],
\end{equation}
we obtain the compact relation
\begin{equation}
  I_1(\theta)=\frac{4}{3}\mathcal{K}_{\rm fs}(\theta).
\end{equation}
Substituting this and $\int_0^1 d\mu\,(1+\mu^2)=4/3$ back into the linewidth
expression yields
\begin{equation}
\begin{aligned}
  \Gamma_\xi^{(\mathrm{fs})}
  &\approx
  \frac{\pi^2\xi^2}{(N+1)^3}\frac{3\gamma}{16}
  \left[
    \frac{4}{3}
    +(-1)^{N+\xi}\frac{4}{3}\mathcal{K}_{\rm fs}(\theta_{N+1})
  \right] \\
  &=
  \frac{\pi^2\xi^2}{(N+1)^3}\,
  \frac{\gamma}{4}
  \Big[1+(-1)^{N+\xi}\mathcal{K}_{\rm fs}(\theta_{N+1})\Big].
\end{aligned}
\end{equation}
This gives the analytic free-space contribution quoted in
Eq.~\eqref{eq:Gamma-fs-analytic}.

\section{Discrete autocorrelation and the closed form of $\mathcal{C}_\xi(\Delta)$}
\label{app:S}

In this appendix, we derive a discrete autocorrelation function used to calculate the collective decay rate shown in Eq. \eqref{eq:Gammafs_main} and energy shift shown in Eq. \eqref{eq:Jfs_main}. 

For the Bragg-edge Dirichlet mode, both Eqs. \eqref{eq:Gammafs_main} and \eqref{eq:Jfs_main} have the mathematical form 
\begin{equation}
F_\xi:=\sum_{j,l=1}^{N}c_\xi^{*}(j)c_\xi(l) f(\beta|j-l|)
\end{equation}
where 
$f(\beta|j-l|)=\mathcal{K}_{\rm fs}(k_0 z_{jl})$ for $\Gamma_\xi^{(\mathrm{fs})}$ and $f(\beta|j-l|)=\mathcal{L}_{\rm fs}(k_0 z_{jl})$ for $J_\xi^{(\mathrm{fs})}$. 
We can define the autocorrelation function
\begin{equation}
  \mathcal{C}_\xi(\Delta)
  \equiv
  \sum_{n=1}^{N-\Delta}c_\xi^{*}(n+\Delta)c_\xi(n),
  \qquad
  \Delta=0,1,\dots,N-1.
\end{equation}
Since $f(\beta\Delta)$ depends only on $\Delta=|j-l|$, grouping the double sum by the distance gives
\begin{equation}
  \sum_{j,l=1}^{N}c_\xi^{*}(j)c_\xi(l) f(\beta|j-l|)
  =f(0)\mathcal{C}_\xi(0)+2\sum_{\Delta=1}^{N-1}f(\beta\Delta)\mathcal{C}_\xi(\Delta).
\end{equation}

To obtain the closed form of $\mathcal{C}_\xi(\Delta)$, we insert the explicit mode profile $ c_\xi(j)$ shown in Eq. \eqref{eq:Dirichlet_main} which yields
\begin{equation}
\begin{aligned}
  \mathcal{C}_\xi(\Delta)
  &=\frac{2}{N+1}(-1)^\Delta
    \sum_{n=1}^{N-\Delta}\sin\big(a(n+\Delta)\big)\sin(an) \\
  &=\frac{(-1)^\Delta}{N+1}
    \sum_{n=1}^{N-\Delta}
    \Big[\cos(a\Delta)-\cos(2an+a\Delta)\Big] \\
    &=\frac{(-1)^\Delta}{N+1}
   \left[(N-\Delta)\cos(a\Delta)-\sum_{n=1}^{N-\Delta}\cos(2an+a\Delta)\right].
\end{aligned}
\end{equation}
The remaining cosine sum is evaluated with
\begin{equation}
  \sum_{n=1}^{M}\cos(n\theta+\varphi)
  =\frac{\sin(M\theta/2)}{\sin(\theta/2)}
   \cos\!\left(\frac{(M+1)\theta}{2}+\varphi\right),
\end{equation}
where $M=N-\Delta$, $\theta=2a$, and $\varphi=a\Delta$. Since $(N+1)a=\pi\xi$, one finds
\begin{equation}
  \sum_{n=1}^{N-\Delta}\cos(2an+a\Delta)
  =-\frac{\sin((\Delta+1)a)}{\sin a}.
\end{equation}
Therefore
\begin{equation}
  \mathcal{C}_\xi(\Delta)
  =\frac{(-1)^\Delta}{N+1}
   \left[(N-\Delta)\cos(a\Delta)+\frac{\sin((\Delta+1)a)}{\sin a}\right].
\end{equation}
Using
\begin{equation}
  \frac{\sin((\Delta+1)a)}{\sin a}
  =\cos(a\Delta)+\cot a\,\sin(a\Delta),
\end{equation}
we finally obtain
\begin{equation}
  \mathcal{C}_\xi(\Delta)
  =\frac{(-1)^\Delta}{N+1}
   \Big[(N+1-\Delta)\cos(a\Delta)+\cot a\,\sin(a\Delta)\Big].
\end{equation}

\section{Free-space contribution to the collective energy shift}
\label{app:Jfs}

This appendix derives the free-space contribution to the real part of the
most subradiant eigenvalues within the same Bragg-edge framework used for
the linewidth.

\subsection{Shift kernel and single-sum representation}

Starting from Eq.~\eqref{eq:Jfs_main}, we define the free-space shift kernel
through
\begin{equation}
  2\,\Im\!\left[V(r)e^{ik_0r}\right]
  \equiv
  \gamma\,\mathcal{L}_{\mathrm{fs}}(k_0r).
\end{equation}
Let $x=k_0r$. Using
\begin{equation}
  V(r)=\frac{3\gamma}{4}
  \left[
    -\frac{i}{x}
    +\frac{1}{x^2}
    +\frac{i}{x^3}
  \right],
  \qquad
  e^{ix}=\cos x+i\sin x,
\end{equation}
we obtain
\begin{equation}
\begin{aligned}
  -\frac{i}{x}e^{ix}
  &= \frac{\sin x}{x}-i\frac{\cos x}{x},\\
  \frac{1}{x^2}e^{ix}
  &= \frac{\cos x}{x^2}+i\frac{\sin x}{x^2},\\
  \frac{i}{x^3}e^{ix}
  &= -\frac{\sin x}{x^3}+i\frac{\cos x}{x^3}.
\end{aligned}
\end{equation}
Therefore,
\begin{equation}
\begin{aligned}
  V(r)e^{ik_0r}
  =&\frac{3\gamma}{4}\left[
      \frac{\sin x}{x}
      +\frac{\cos x}{x^2}
      -\frac{\sin x}{x^3}
    \right]\\
   &+i\frac{3\gamma}{4}\left[
      -\frac{\cos x}{x}
      +\frac{\sin x}{x^2}
      +\frac{\cos x}{x^3}
    \right],
\end{aligned}
\end{equation}
which gives
\begin{equation}
  \mathcal{L}_{\mathrm{fs}}(x)
  =
  \frac{3}{2}
  \left[
    -\frac{\cos x}{x}
    +\frac{\sin x}{x^2}
    +\frac{\cos x}{x^3}
  \right].
  \label{eq:Lfs-app}
\end{equation}

For the geometry-dependent collective energy shift, the diagonal self-energy
(single-atom Lamb shift) is absorbed into the renormalized bare transition
frequency. Hence only the off-diagonal terms are kept in the free-space part.
Using the distance-grouping identity derived in Appendix~\ref{app:S}, we obtain
\begin{equation}
  J_\xi^{(\mathrm{fs})}
  =
  \gamma\sum_{\Delta=1}^{N-1}
  \mathcal{L}_{\mathrm{fs}}(\beta\Delta)\,
  \mathcal{C}_\xi(\Delta),
  \qquad
  \beta\equiv k_0d,
  \label{eq:Jfs-single-app}
\end{equation}
where
\begin{equation}
  \mathcal{C}_\xi(\Delta)
  =
  \frac{(-1)^\Delta}{N+1}
  \Big[
    (N+1-\Delta)\cos(a\Delta)+\cot a\,\sin(a\Delta)
  \Big].
  \label{eq:CDelta-exact-appJ}
\end{equation}
Equation~\eqref{eq:Jfs-single-app} is the exact finite-\(N\) free-space
contribution within the Bragg-edge ansatz.

\subsection{Deep-subwavelength asymptotics for most subradiant modes}

For a fixed Bragg-edge mode with $\xi\ll N$, one has
\begin{equation}
  a=\frac{\pi\xi}{N+1}\ll1.
\end{equation}
Expanding Eq.~\eqref{eq:CDelta-exact-appJ} for fixed $\Delta$ at large $N$
gives
\begin{equation}
  \mathcal{C}_\xi(\Delta)
  =
  (-1)^\Delta
  \left[
    1-\frac{a^2\Delta^2}{2}
    +O\!\left(\frac{a^2\Delta^3}{N+1}\right)
    +O(a^4\Delta^4)
  \right].
  \label{eq:SDelta-small-a-appJ}
\end{equation}
Likewise, for $x\ll1$, the shift kernel has the expansion
\begin{equation}
  \mathcal{L}_{\mathrm{fs}}(x)
  =
  \frac{3}{2}
  \left[
    \frac{1}{x^3}
    -\frac{1}{2x}
    +\frac{3x}{8}
    +O(x^3)
  \right].
  \label{eq:Lfs-small-x-app}
\end{equation}

Substituting Eqs.~\eqref{eq:SDelta-small-a-appJ} and
\eqref{eq:Lfs-small-x-app} into Eq.~\eqref{eq:Jfs-single-app}, we obtain
\begin{equation}
\begin{aligned}
  J_\xi^{(\mathrm{fs})}(N,d)
  \approx\;
  &\frac{3\gamma}{2\beta^3}
  \sum_{\Delta=1}^{N-1}\frac{(-1)^\Delta}{\Delta^3}
  -\frac{3\gamma}{4\beta}
  \sum_{\Delta=1}^{N-1}\frac{(-1)^\Delta}{\Delta}
\\
  &-\frac{3\gamma a^2}{4\beta^3}
  \sum_{\Delta=1}^{N-1}\frac{(-1)^\Delta}{\Delta}
  +\cdots .
\end{aligned}
\label{eq:Jfs-expand-app}
\end{equation}

Retaining the $\xi$-independent leading terms and the leading $\xi$-dependent
finite-size correction, we obtain
\begin{equation}
  \begin{aligned}
  J_\xi^{(\mathrm{fs})}(N,d)
  \approx\;&
  \frac{3\gamma}{2\beta^3}\sum_{\Delta=1}^{N-1}\frac{(-1)^\Delta}{\Delta^3}
  -\frac{3\gamma}{4\beta}\sum_{\Delta=1}^{N-1}\frac{(-1)^\Delta}{\Delta} \\
  &-\frac{3\gamma a^2}{4\beta^3}
  \sum_{\Delta=1}^{N-1}\frac{(-1)^\Delta}{\Delta}
  +O\!\left(\gamma\beta\right)
  +O\!\left(\gamma\frac{a^2}{\beta}\right).
  \end{aligned}
  \label{eq:Jfs-asym-pre}
\end{equation}

For large $N$, the sums may be extended to infinity and we have
\begin{equation}
  \sum_{\Delta=1}^{\infty}\frac{(-1)^\Delta}{\Delta^3}
  =-\frac{3}{4}\zeta(3),
  \qquad
  \sum_{\Delta=1}^{\infty}\frac{(-1)^\Delta}{\Delta}
  =-\ln2.
\end{equation}
Therefore, Eq.~\eqref{eq:Jfs-asym-pre} becomes
\begin{equation}
\begin{aligned}
  J_\xi^{(\mathrm{fs})}(N,d)
  \approx\;
  &-\frac{9\gamma}{8}\frac{\zeta(3)}{\beta^3}
  +\frac{3\gamma\ln2}{4}\frac{1}{\beta}
  +\frac{3\gamma a^2\ln2}{4}\frac{1}{\beta^3}
  +\cdots
\\
  =\;
  &-\frac{9\gamma}{8}\frac{\zeta(3)}{(k_0d)^3}
  +\frac{3\gamma\ln2}{4}\frac{1}{k_0d}\\
  &+\frac{3\gamma\pi^2\xi^2\ln2}{4}\,
   \frac{1}{(N+1)^2}\frac{1}{(k_0d)^3}
  +\cdots,
\end{aligned}
\label{eq:Jfs-asymp-general-app}
\end{equation}
which is Eq.~\eqref{eq:Jfs-finite-main} in the main text.

It is convenient to define the thermodynamic-limit free-space energy shift
\begin{equation}
  J_{\infty}^{(\mathrm{fs})}(d)
  \approx
  -\frac{9\gamma}{8}\frac{\zeta(3)}{(k_0d)^3}
  +\frac{3\gamma\ln2}{4}\frac{1}{k_0d}.
  \label{eq:Jinf-fs-app}
\end{equation}
Then Eq.~\eqref{eq:Jfs-asymp-general-app} can be written as
\begin{equation}
  J_\xi^{(\mathrm{fs})}(N,d)
  \approx
  J_{\infty}^{(\mathrm{fs})}(d)
  +
  \frac{3\gamma\pi^2\xi^2\ln2}{4}\,
  \frac{1}{(N+1)^2}\frac{1}{(k_0d)^3}
  +\cdots .
  \label{eq:Jfs-final-app}
\end{equation}

Equation~\eqref{eq:Jfs-final-app} shows that the leading deep-subwavelength
free-space shift is independent of $\xi$ for the Bragg-edge modes,
whereas the first finite-size correction scales as
$\xi^2/(N+1)^2$. This is the free-space origin of the finite-size behavior of
the real part discussed in the main text.

%


\end{document}